\documentclass[twocolumn, nofootinbib, showpacs, superscriptaddress]{revtex4-1} 
\usepackage{amsfonts}
\usepackage{ulem}
\usepackage{amsmath}
\usepackage{dsfont}
\usepackage[pdftex]{graphicx}
\usepackage{epstopdf}
\usepackage{ulem}
\usepackage{textcomp} 
\usepackage{wasysym} 
\usepackage{placeins}
\usepackage{graphicx}
\usepackage{bm}
\usepackage[varg]{txfonts} 
\usepackage{esint}
\usepackage{xcolor}
\usepackage{cancel}
\usepackage{comment}
\usepackage{hyperref}
\hypersetup{
    colorlinks=true,
    linkcolor=red,
    citecolor=blue,
} 
\specialcomment{supplement}{\begingroup\color{cyan}\footnotesize}{\endgroup}
\specialcomment{formula}{\begingroup\footnotesize}{\endgroup}
\excludecomment{details}

\renewcommand{\H}[0]{{\rm{H}}}
\newcommand{\W}[0]{{\rm{W}}}
\newcommand{\sigp}{{\sigma_{\! \! p}}}
\newcommand{\sigx}{{\sigma_{\! \! x}}}

\renewcommand{\d}[0]{{\rm{d}}}
\newcommand{\del}[0]{\partial }

\newcommand{\vol}[2]{\hspace{-0.8mm}\mbox{$\text{d}^{\hspace{-0.0mm}#1}$}\hspace{-0.2mm}#2\hspace{0.8mm}\ }
\renewcommand{\v}[1]{\bm{#1} }
\newcommand{\vx}[0]{\bm{x} }

\newcommand{\vp}[0]{\bm{p} }
\newcommand{\vnabla}[0]{\bm{\nabla} }
\begin{document}
\title{Schr\"odinger method as N-body double and UV completion of dust} 
\author{Cora Uhlemann} 
\email{cora.uhlemann@physik.lmu.de}
\affiliation{Arnold Sommerfeld Center for Theoretical Physics, Ludwig-Maximilians-Universit\"at, Theresienstr.\,37, 80333 Munich, Germany} 
\affiliation{Excellence Cluster Universe, Boltzmannstr.\,2, 85748 Garching, Germany} 
\author{Michael Kopp} 
\email{michael.kopp@physik.lmu.de}
\affiliation{Arnold Sommerfeld Center for Theoretical Physics, Ludwig-Maximilians-Universit\"at, Theresienstr.\,37, 80333 Munich, Germany} 
\affiliation{Excellence Cluster Universe, Boltzmannstr.\,2, 85748 Garching, Germany} 
\affiliation{University Observatory, Ludwig-Maximilians University, Scheinerstr.\,1, 81679 Munich, Germany} 
\author{Thomas Haugg} 
\email{thomas.haugg@physik.lmu.de}
\affiliation{Arnold Sommerfeld Center for Theoretical Physics, Ludwig-Maximilians-Universit\"at, Theresienstr.\,37, 80333 Munich, Germany} 

\begin{abstract}
We investigate large-scale structure formation of collisionless dark matter in the phase space description based on the Vlasov (or collisionless Boltzmann) equation whose nonlinearity is induced solely by gravitational interaction according to the Poisson equation. Determining the time-evolution of density and peculiar velocity demands solving the full Vlasov hierarchy for the moments of the phase space distribution function. In the presence of long-range interaction no consistent truncation of the hierarchy is known apart from the pressureless fluid (dust) model which is incapable of describing virialization due to the occurrence of shell-crossing singularities and the inability to generate vorticity and higher cumulants like velocity dispersion. Our goal is to find a simple ansatz for the phase space distribution function that approximates the full Vlasov distribution function without pathologies in a controlled way  and therefore can serve as theoretical N-body double and as a replacement for the dust model. We argue that the coarse-grained Wigner probability distribution obtained from a wave function fulfilling the Schr\"odinger-Poisson equation (SPE) is the sought-after function. We show that its evolution equation approximates the Vlasov equation and therefore also the dust fluid equations before shell-crossing, but cures the shell-crossing singularities and is able to describe regions of multi-streaming and virialization. This feature was already employed in cosmological simulations of large-scale structure formation by Widrow \& Kaiser (1993). 
The coarse-grained Wigner ansatz allows to calculate all higher moments from density and velocity analytically, thereby incorporating nonzero higher cumulants in a self-consistent manner. On this basis we are able to show that the Schr\"odinger method (ScM) automatically closes the corresponding hierarchy such that it suffices to solve the SPE in order to directly determine density and velocity and all higher cumulants.
\end{abstract}

\maketitle
\section{Introduction}

The standard model of large-scale structure (LSS) formation and halo formation is based on collisionless cold dark matter (CDM), a yet unknown particle species that for  purposes of LSS and larger halos can be assumed to interact only gravitationally and to be cold or initially single-streaming. We are therefore interested in the dynamics of a large collection of identical point particles that via gravitational instability evolve from initially small density perturbations into eventually bound structures, like halos that are distributed along the loosely bound LSS composed of superclusters, sheets, and filaments \cite{P80,SWF05,TSM13}. All these structures depend on cosmological parameters, in particular the background energy density of CDM and the cosmological constant. We therefore require accurate modelling and theoretical understanding of CDM dynamics to extract those cosmological parameters from observations. While the shape of the LSS can be reasonably well described by modelling the CDM as a pressureless fluid (dust), it necessarily fails at small scales where multiple streams form. Multi-streaming is especially important for halo formation -- virialization, but already affects LSS and its observation in redshift-space.

On sub-Hubble scales and for non-relativistic velocities the Newtonian limit of the Einstein equations is sufficient to describe the time evolution of structures within the universe \cite{CZ11,GW12,KUH14}. Furthermore the large number of particles under consideration suppresses collisions such that the phase space dynamics is only affected by the smooth Newtonian potential \cite{G68}. Therefore the time-evolution of the phase space distribution function $f(t,\v{x},\v{p})$ is governed by the Vlasov (or collisionless Boltzmann) equation whose nonlinearity is induced by the gravitational force obtained from the Poisson equation sourced by $\int \vol{3}{p}\!f(t,\v{x},\v{p})$. 

Even though this model seems to be quite simple from a conceptual point of view, no general solution is known and one usually has to resort to N-body simulations which tackle the problem of solving the dynamical equations numerically, see \cite{T02,SWF05,SWF06,SW09,AHK12,HAK13}. From the analytical point of view, different methods to describe LSS formation based on the dust model have been developed. The dust model describes CDM as a pressureless fluid using hydrodynamic equations \cite{P80}, and is studied especially in the context of perturbation theory. Among them the two most commonly used methods are the Eulerian framework describing the dynamics of density and velocity fields, see \cite{B02}, and the Lagrangian description following the field of trajectories of particles \cite{B94}. The dust model is an exact solution to the Vlasov equation which describes absolutely cold dark matter and works quite well in the linear and quasi-linear regime of LSS formation. But the dust model not only fails to catch the dynamics when multiple streams occur in the N-body dynamics, but actually runs into so called shell-crossing singularities or caustics forming at the smallest scales. One might therefore say that the dust model is UV-incomplete.\\

A possibility to circumvent the formation of singularities and to restore agreement with simulations in the weakly nonlinear regime is to introduce an artificial viscosity term in the pressureless fluid equations which is effective only in regions where the dust evolution would predict a singularity. This phenomenological model proposed in \cite{G89} is known as adhesion approximation and was shown to be able to reproduce the skeleton of the cosmic web in \cite{WG90}. However, such ad-hoc constructions remain quite unsatisfying from a conceptual point of view; for example the size of formed structures directly depends on the viscosity parameter rather then the initial conditions and it is unclear how well the Vlasov equation is approximated. 

A more general reasoning was pursued in the direction of coarse-grained perturbation theory which led to models that were argued to incorporate adhesive features. When the dynamical evolution of a many-body system is described by means of a continuous phase space distribution one has to consider coarse-grained or macroscopic quantities, thereby neglecting detailed information about the microscopic degrees of freedom. Although at a first glance this might seem inconvenient, it is indeed an advantageous point of view, especially when comparing to data inferred from observations or simulations, that are fundamentally coarse-grained. Therefore the dynamical evolution of smoothed density and velocity fields relevant for cosmological structure formation has been under investigation, see for example \cite{D00,BD05}, where it was argued that coarse-graining may lead automatically to adhesive behavior. Furthermore it was shown in \cite{P12} that for averaged fields the correspondence between the occurence of velocity dispersion and multi-streaming phenomena due to shell-crossing breaks down. This is due to the fact that the coarse-graining introduces a nonzero velocity dispersion between the particles within each coarse-graining cell which mimics microscopic velocity dispersion connected to genuine multi-streaming.\\

Solving the Vlasov equation is equivalent to solving the infinite coupled hierarchy of equations for the cumulants of the distribution function $f$ with respect to momentum $\v{p}$. This means that in order to determine the time evolution of the zeroth and first cumulants, related to density and velocity, all higher cumulants starting with velocity dispersion are relevant, see \cite{PS09}. Only neglecting them entirely is consistent \cite{PS09}; in this case one is lead to the popular dust model \cite{P80}. Gravity is the dominant force on cosmological scales and in the early stages of gravitational instability matter is distributed very smoothly with nearly single-valued velocities. Therefore the dust model has proven quite successful in describing the evolution as long as the collective motion of particles is well-described by this coherent flow. However, as soon as the density contrast becomes non-linear, multiple streams form and become relevant in the Vlasov dynamics while caustics -- called `shell-crossing' singularities -- are developed indicating that the underlying approximations are no longer justified and the model looses its predictability. The problem of developing singularities and failure of being a good description afterwards, also occurs in the first order Lagrangian solution, called Zel'dovich approximation \cite{Z70}, which is the exact solution in the plane-parallel collapse studied in Sec.\,\ref{sec:numerics}.

The Schr\"odinger method (ScM), originally proposed in \cite{WK93,DW96} as numerical technique for following the evolution of CDM, models CDM as a complex scalar field obeying the coupled Schr\"odinger-Poisson equations (SPE) \cite{SC02,SC06,G95} in which $\hbar$ merely is a free parameter that can be chosen at will and determines the phase space resolution. The ScM is comprised of two parts; (1) solving the SPE with desired initial conditions and (2) taking the Husimi transform \cite{H40} to construct a phase space distribution from the wave function. The correspondence between distribution functions in classical mechanics and phase space representations of quantum mechanics has been investigated in detail by \cite{T89}, both analytically as well as by means of numerical examples. It turned out that the Wigner function, obtained from a wave function fulfilling the SPE, corresponds poorly to classical dynamics. In contrast, the coarse-grained Wigner or Husimi distribution was shown to be indeed a good model for coarse-grained classical mechanics \cite{T89, WK93}. \\

The SPE can be seen as the non-relativistic limit of the Klein-Gordon-Einstein equations \cite{W97, GG12}. From this perspective the physical interpretation (if $\hbar$ takes the value of the Planck constant) is that CDM is actually a non-interacting and non-relativistic Bose-Einstein condensate in which case the SPE can be interpreted as a special Gross-Pitaevskii equation, see \cite{R13} for a review.  In plasma and solid state physics as well as mathematical physics the equation is known as Choquard equation \cite{L77, AS01}.  In the context of gravitational state reduction this equation, denoted by Schr\"odinger-Newton equation, was studied e.g. in \cite{MPT98}. There have also been investigations on the connection between general fluid dynamics and wave mechanics \cite{M27, S80}.

The similarity between the SPE and the dust model has been also employed in the context of wave mechanics. There the so-called free-particle approximation (based on the free-particle Schr\"odinger equation, see \cite{Th11}) was shown to closely resemble the Zel'dovich approximation \cite{SC02, SC06} while avoiding singularities. In some works a modified SPE system with an added quantum pressure term was considered, \cite{JLH09,TWJ11} which then is equivalent to the usual fluid system. Clearly this approach is not advantageous since the fluid description is known to break down at shell-crossing. This had lead to the claim in \cite{JLH09} that also the Schr\"odinger method breaks down. In \cite{SK02} perturbation theory based on the SPE in the  limit $\hbar \rightarrow 0$ was considered where it was emphasized that shell-crossing singularities are avoided. However their calculations assumed $\hbar =0$ identically, which leads to results equivalent to standard perturbation theory (SPT) based on a dust fluid, without solving the shell-crossing problem. 

\begin{center}
\begin{figure*}[!]
\includegraphics[width=0.90\textwidth]{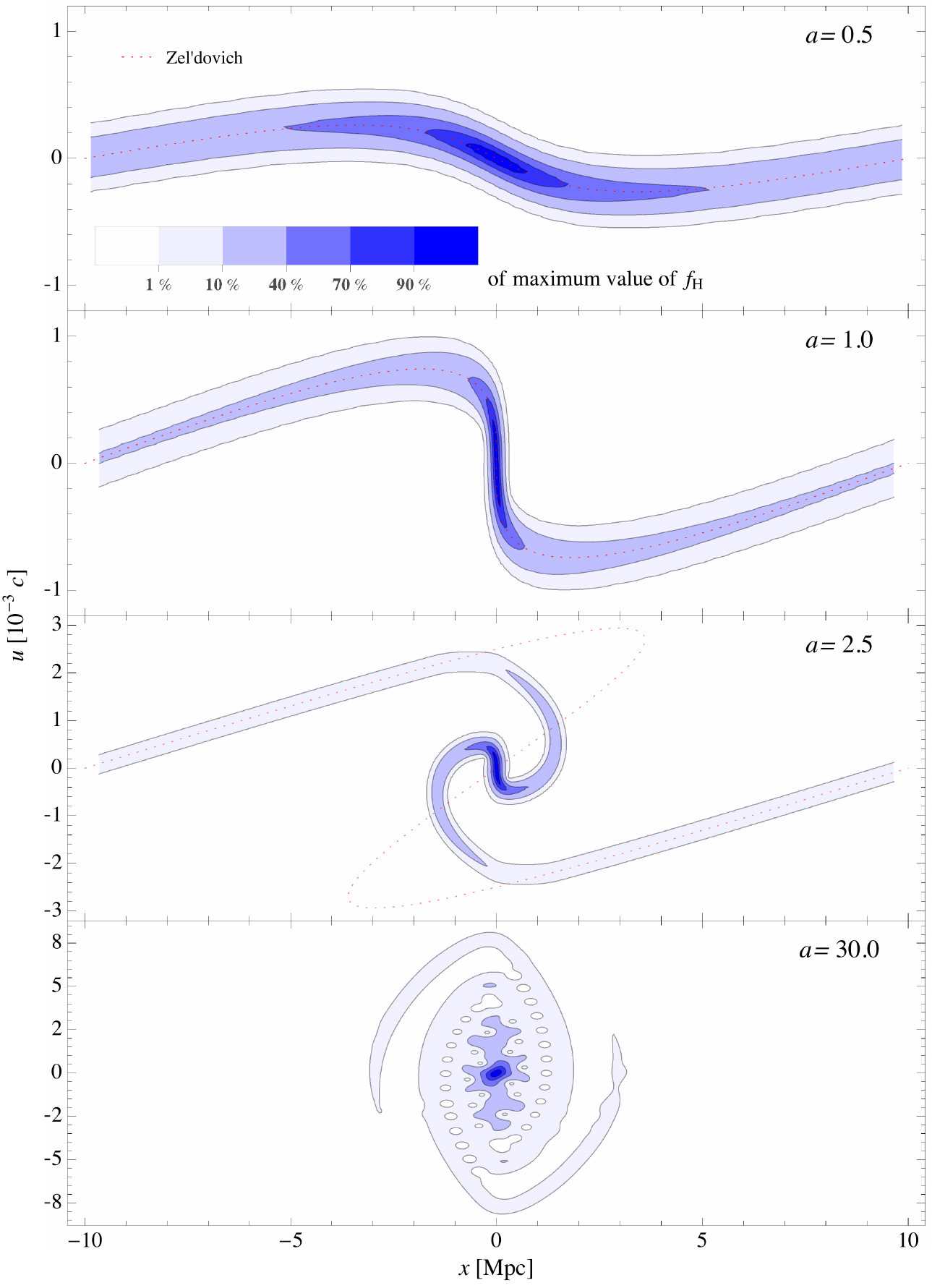}\\
\caption{Collapse of a pancake (plane-parallel) density profile on a Einstein-de Sitter background as seen in phase space using the ScM. \textit{blue} contours: Phase space density $f_{\rm H}$ calculated from Eqs.\,(\ref{schrPoissEqFRW}, \ref{Husimi}) at four moments in time.  \textit{red dotted} line: the Zel'dovich solution of Eq.\,\eqref{ZeldoPancake} is the exact dust solution, valid until $a=1$. Only the first panel of the four characteristic moments can be described by dust. Shell-crossing (2nd panel), multi-streaming (3rd panel) and viralisation (4th panel) are accessible with the ScM but not with dust. That the dynamics corresponds to CDM is proven in Sec.\,\ref{subsection:husimivlasovmap}. How to obtain cumulants without constructing $f_{\rm H}$ is shown in Sec.\,\ref{HusHierarchy}.}
\label{fig:phaseplot}
\end{figure*}
\end{center}

That the ScM is a viable model for cosmological structure formation and in particular capable of describing multi-streaming was exemplarily demonstrated by means of numerical examples in \cite{WK93,DW96, SBRB13}. However, the bulk of these investigations were aimed at replacing N-body simulations by a numerical solution to the SPE. Therefore the methods applied therein are unsuitable and inconvenient for the genuine analytical approach we want to establish. In \cite{WK93,DW96} a superposition of $N$ Gaussian wave packets was used as initial wave function, thereby closely resembling the $N$ particles in a N-body simulation. In \cite{SBRB13} CDM was modeled by $N$ wave functions coupled via the Poisson equation. We will study the case of a single wave function on an expanding background with nearly cold initial conditions. The result suggests that indeed the ScM is a substantially better suited analytical tool to study CDM dynamics than the dust model: in the single-stream regime they stay arbitrarily close to each other, but while dust fails and stops when multi-streaming should occur, the Schr\"odinger wave function continues without any pathologies and behaves like multi-streaming CDM when interpreted in a coarse-grained sense. Although it was already observed in \cite{SC02} that the wave function does not run into singularities, it was claimed that it still cannot describe multi-streaming or virialization. Indeed, our numerical example closely resembling that of \cite{SC02}, but generalised to an expanding background, proves the contrary. Fig\,\ref{fig:phaseplot} shows the dynamics of the Husimi function $f_{\rm H}$ using the ScM: the density remains finite at shell-crossing, $f_{\rm H}$ forms multi-stream regions and ultimately virializes. None of these features necessary for a full description of LSS and halo formation are accessible with the dust model.

\paragraph*{Goal}
The aim of this paper is to present the Schr\"odinger method, already investigated in the context of cosmological simulations, as a theoretical N-body double for the phase space distribution function $f$. We show that phase space density $f_{\rm H}$ obtained from the ScM solves the Vlasov equation approximately but in a controlled manner. We demonstrate that $f_{\rm H}$ closes the hierarchy of moments automatically but yet allows for multi-streaming and virialization. We give explicit analytic expressions for higher order non-vanishing cumulants, like velocity dispersion, in terms of the wave function and in terms of the macroscopic physical density and velocity fields. This constitutes a new approach to tackle the closure problem of the Vlasov hierarchy apart from truncation or restricting oneself to the dust model and its limitations. We shed light on the physical interpretation by means of a numerical study of pancake formation. In summary this means that the ScM models CDM in a well-behaved manner with initial conditions and single-stream dynamics arbitrarily close to dust. Unlike dust, the ScM captures all relevant physics for describing CDM dynamics even in the deeply nonlinear regime and does not fail on the smallest scales, therefore providing a UV-completion of dust.

\paragraph*{Structure}
This paper is organized as follows: In Sec.\,\ref{sec:PS-CDM} we review the phase space description of cold dark matter and explain how one is lead to the Vlasov equation on an expanding background. After  introducing the dust model we re-derive the coarse-grained Vlasov equation. We then introduce the Wigner function as an ansatz for the phase space distribution and explain its connection to the dust model. We derive the corresponding Wigner-Vlasov equation as well as its coarse-grained version and discuss their relations to the Vlasov equation and the coarse-grained Vlasov equation, respectively. In Sec.\,\ref{sec:Hierarchy} we determine the moments of the three different phase space distributions  -- the dust model, the Wigner function and the coarse-grained Wigner or Husimi distribution. In Sec.\,\ref{sec:numerics} we investigate the pancake collapse to illustrate that the dynamics of the complex scalar field is free from the pathologies of the dust fluid and serves therefore both as a theoretical N-body double and as a UV completion of dust. On this basis we explain how the closure of the hierarchy of moments can be achieved and finally discuss the implications. In Sec.\,\ref{sec:prospects} we make suggestions about possible future research based on ScM and conclude in Sec.\,\ref{sec:conclusion}.
\vspace{-1cm}

\section{Phase-space description of cold dark matter}
\label{sec:PS-CDM}
\subsection{From Klimontovich to Vlasov equation}
The exact one-particle (Klimontovich) phase space density $f_{\rm K}$ of $N$ identical particles following trajectories $\{\vx_i(t), \vp_i(t)\}$, $i \in {1,...,N}$, 
in phase space is given by a sum of $\delta$-functions
\begin{align}
f_{\rm K} (t,\vx,\vp) = \frac{1}{N}\sum_{i=1}^N \delta_{\rm D}\left(\vx-\vx_i(t)\right)\delta_{\rm D}\left(\vp-\vp_i(t)\right) \,.
\end{align}
We use comoving coordinates $\vx$ with associated conjugate momentum $\vp=a^2m\ d\vx/dt$, where $a$ is the scale factor satisfying the Friedmann equation of a $\Lambda$CDM or Einstein-de Sitter universe.\footnote{More generally, any expansion history is allowed as long as metric perturbations are only sourced by CDM.} For convenience we will in general suppress the $t$-dependence of the distribution function in the following. This phase space density obeys the Klimontovich equation \cite{K69} encoding phase space density conservation along phase space trajectories
\begin{align}
\frac{Df_{\rm K}}{dt} = \frac{\partial f_{\rm K}}{\partial t} + \frac{d\vx}{dt}\cdot\frac{\del f_{\rm K}}{\del \vx} + \frac{d\vp}{dt}\cdot\frac{\del f_{\rm K}}{\del \vp}  = 0 \,.
\end{align}
Upon using the equations of motion for non-relativistic particles with trajectories $\{\vx_i(t), \vp_i(t)\}$ one arrives at
\begin{subequations}
\label{KlimoPoissonEq}
\begin{align}
\label{KlimoEq}
\partial_t f_{\rm K}&=   -\frac{\vp}{a^2 m}\cdot \vnabla_{\! \!  x} f_{\rm K} + m \vnabla_{\! \!  x} V \cdot \vnabla_{\! \!  p} f_{\rm K} \,.
\end{align}
The nonlinearity in \eqref{KlimoEq} is induced by the fact that the Newtonian potential $V$ describes gravitational interaction and therefore depends through the Poisson equation on the density field given by the integral of the distribution function over momentum\\
\begin{align}
\label{PoissonEq}
\Delta V &= \frac{4\pi G\rho_0}{a} \left(\int \vol{3}{p} \!f_{\rm K} - 1 \right) \,,
\end{align}
\end{subequations}
where $\rho_0$ is the (constant) comoving matter background density such that $f_{\rm K}$ has a background value  or spatial average value $\langle \int \vol{3}{p}\!f_{\rm K}\rangle_{\rm vol}=1$. When symbols like $\vnabla$ or $\Delta= \vnabla\cdot\vnabla$ are used without subscripts they refer to spatial derivatives $\vnabla_{\! \!  x}$ or $\Delta_{x}$, respectively.

Retaining all details concerning the microstate of a system, the spiky Klimontovich density is not really of practical use. Rather one is interested in the statistical average taken over an ensemble of different realizations of the distribution of the $N$ particles. This information is contained within the smooth one-particle phase space density $f_1$ given by
\begin{align}
f_1(t,\vx,\vp) = \langle f_{\rm K}(t,\vx,\vp) \rangle \,,
\end{align}
where angle brackets denote the ensemble average of microstates $f_{\rm K}$ that lead to the same coarse-grained phase space density. If $V$ was a specified external potential, $f_1$ would obey the same equation as $f_{\rm K}$. However, since $V$ is the gravitational potential computed self-consistently from the particles via \eqref{PoissonEq}, the $\vnabla_{\! \!  x} V \cdot \vnabla_{\! \!  p} f_{\rm K}$ term in \eqref{KlimoEq} is quadratic in $f_{\rm K}$. Therefore when taking the ensemble average to derive an equation for the one-particle distribution function $f_1$ an additional correlation term emerges which involves the irreducible part $f_{2c}$ of two-particle distribution function $f_2(\vx,\vp,\vx',\vp')=f_1(\vx,\vp)f_1(\vx',\vp') + f_{2c}(\vx,\vp,\vx',\vp')$, compare \cite{B93}
\begin{align}
\label{BBGKYEq}
\partial_t f_1&=   -\frac{\vp}{a^2 m}\cdot\vnabla_{\! \!  x} f_1 + m \vnabla_{\! \!  x} V \cdot\vnabla_{\! \!  p} f_1\\
\notag &\quad + m \int \vol{3}{x'}\vol{3}{p'} \vnabla_{\! \!  x} V(\vx-\vx') \cdot  \vnabla_{\! \!  p} f_{2c}(\vx,\vp,\vx',\vp')\,.
\end{align}
This leads to a set of coupled kinetic equations where the $n$-particle distribution in turn depends on the $(n+1)$-particle distribution. This is the so-called BBGKY (Bogoliubov-Born-Green-Kirkwood-Yvon) hierarchy, describing the dynamics of an interacting N-particle system. The resulting equation \eqref{BBGKYEq} for $f_1$ differs from the Klimontovich equation \eqref{KlimoPoissonEq} by a correlation term which vanishes in the absence of pair correlations. Fortunately, for the case of interest here - CDM particles - these collisional effects are completely negligible since they are suppressed by $1/N$ where $N$ is the number of particles, see \cite{G68}. The corresponding Vlasov-Poisson system for the one-particle phase space density $f_1$, which we will denote simply by $f$ from now on, describes collisionless dark matter in the absence of two-body correlations
\begin{subequations}
\label{VlasovPoissonEq}
\begin{align}
\label{VlasovEq}
\partial_t f&=   -\frac{\vp}{a^2 m}\cdot\vnabla_{\! \!  x} f + m \vnabla_{\! \!  x} V \cdot\vnabla_{\! \!  p} f \,, \\
&=\left[ \frac{\vp^2}{2a^2m}+m V(\vx)\right] \left( \overleftarrow{\vnabla}_{\! \!  x} \overrightarrow{\vnabla}_{\! \!  p}- \overleftarrow{\vnabla}_{\! \!  p} \overrightarrow{\vnabla}_{\! \!  x}\right) f \,,\\
\Delta V &= \frac{4\pi G\,\rho_0}{a}  \left(\int \vol{3}{p}\!\!f - 1\ \right) \,.
\end{align}
\end{subequations}
\subsection{Dust model}
The dust model describes CDM as a pressureless fluid with density $n_{\rm d}(\vx)$ and fluid momentum given by an irrotational flow $\vnabla \phi_{\rm d}(\vx)$ which remains single-valued at each point, and therefore absolutely cold, meaning that particle trajectories are not allowed to cross and velocity dispersion cannot arise. This regime is usually referred to as `single-stream', meaning that the validity of this model breaks down as soon as `shell-crossings' occur and multiple streams develop. The corresponding distribution function is given by
\begin{align}
\label{fdust}
f_\d (\vx,\vp)&= n_\d(\vx) \delta_{\mathrm D}\Big(\vp-\vnabla \phi_\d(\vx) \Big) \;.
\end{align}
As we will see in section \ref{sec:Hierarchy}, the Vlasov equation \eqref{VlasovEq} for $f_\d$ implies the hydrodynamical equations for a perfect pressureless fluid with density $n_\d$ and velocity potential $\phi_\d/m$. The fluid equations consist of the continuity equation, the Bernoulli and Poisson equation
\begin{subequations} \label{dustEqs}
\label{fluid}
\begin{align} 
\del_t n_\d &= -\frac{1}{m a^2}\vnabla\cdot \left(n_\d \vnabla \phi_\d\right)\label{conti} \,,\\
\del_t \phi_\d &= -\frac{1}{2 a^2 m} \left(\vnabla\phi_\d\right)^2 -mV_\d \label{euler}\,,\\
\Delta V_\d&=\frac{4\pi G\,\rho_0}{a}\Big(n_\d -  1 \Big) \label{Poissdust}  \,.
\end{align}
\end{subequations}
By defining an irrotational velocity according to $\v{u}_\d=\vnabla\phi_\d/m$ one can rewrite \eqref{conti} and \eqref{euler}  in the following equivalent form
\begin{subequations}\label{Fluideq}
\begin{align}
\del_t n_\d &= -\frac{1}{a^2}\vnabla \cdot(n_\d\v{u}_\d) \,,\\
\del_t \v{u}_\d &= -\frac{1}{a^2}(\v{u}_\d\cdot\vnabla)\v{u}_\d -\vnabla V_\d \,,\\
\vnabla \times \v{u}_\d &= 0 \,. \label{dustconstr}
\end{align}
\end{subequations}

\subsection{Coarse-grained Vlasov equation}
The coarse-grained distribution function $\bar f$ is obtained from $f$ by convolution with a Gaussian of width $\sigx$ and $\sigp$ in $\vx$ and $\vp$ space, respectively. For convenience we will adopt the shorthand operator representation of the smoothing which can be easily obtained by switching to Fourier space 
\begin{align}
\label{cgf}
\notag  \bar f (\vx,\vp)&=\int \frac{\vol{3}{x'}\vol{3}{p'}}{(2 \pi\sigx\sigp)^3 } \exp\left[-\frac{(\vx-\vx')^2}{2\sigx^2}-\frac{(\vp-\vp')^2}{2\sigp^2} \right] f(\vx',\vp')\,,\\
 \bar f &= \exp\left(\frac{\sigx^2}{2}\Delta_x+\frac{\sigp^2}{2}\Delta_p\right) f\,.
\end{align}
The corresponding coarse-grained Vlasov equation as given in \cite{MW03} is easily obtained from the usual Vlasov equation \eqref{VlasovPoissonEq} by applying the smoothing operator. We employ the following identity for the smoothing operator 
\begin{align}
\label{productrule}
\exp(\Delta)(AB)=[\exp(\Delta)A]\exp\left(2\overleftarrow{\vnabla}\overrightarrow{\vnabla}\right)[\exp(\Delta)B] \,,
\end{align}
in order to express the coarse-graining of a product in terms of its coarse-grained factors. The result is the cosmological analogue to the evolution equation for coarse-grained classical distribution 
\vspace{-0.7cm}
\begin{widetext}
\begin{subequations}
\label{cgVlasovEq}
\begin{eqnarray} 
\partial_t  \bar f&&= -\frac{\vp}{a^2 m}\vnabla_{\! \!  x}  \bar f -\frac{\sigp^2}{a^2 m}\vnabla_{\! \!  x}\vnabla_{\! \!  p}  \bar f  +m\vnabla_{\! \!  x} \bar V \exp(\sigx^2\overleftarrow{\vnabla}_{\! \!  x}\overrightarrow{\vnabla}_{\! \!  x})\vnabla_{\! \!  p}  \bar f \label{cgVlasovEqa} \,,\\
&&= \exp \left(\frac{\sigx^2}{2}\Delta_x + \frac{\sigp^2}{2}\Delta_p \right) \left[ \frac{\vp^2}{2a^2m}+m V\right]  \exp \left(\sigx^2 \overleftarrow{\vnabla}_{\! \!  x} \overrightarrow{\vnabla}_{\! \!  x}+ \sigp^2 \overleftarrow{\vnabla}_{\! \!  p} \overrightarrow{\vnabla}_{\! \!  p}\right) \left( \overleftarrow{\vnabla}_{\! \!  x} \overrightarrow{\vnabla}_{\! \!  p}- \overleftarrow{\vnabla}_{\! \!  p} \overrightarrow{\vnabla}_{\! \!  x}\right)  \bar f \label{cgVlasovEqb} \,,
\end{eqnarray}
\end{subequations}
\end{widetext}
which was given in \cite{T89} for $a=1$ and units where $\sigx^2=\sigp^2= \hbar /2$.

Note that this result holds on a FRW background with cosmic time $t$, comoving $\bm x$ and canonical conjugate 1-form $\bm p$, where $\bar V$ fulfills Eq.\,\eqref{PoissonEq} with $f$ is replaced by $ \bar f$. If derivative operators like $\vnabla_{\! \!  x}$ and $\vnabla_{\! \!  p}$ carry left or right arrows over them, they specify that they only act on quantities on their left or right hand side, respectively. The notation of Eq.\,\eqref{cgVlasovEq} is the same as used in \cite{T89}. \\
At a first glance the coarse-graining introduced in \eqref{cgf} might seem like an unfavorable artifact which complicates calculations on the one hand and erases relevant information on the other hand. However, one has to bear in mind that when sampling the distribution function numerically using a finite number of particles, a coarse-graining is inevitable to provide a proper phase space description \cite{BD05}. This is of particular importance since solving the Vlasov-Poisson equation analytically is a formidable task and one typically has to resort to numerical simulations, for example N-body codes  \cite{T02,SWF05,SWF06,SW09,AHK12,HAK13}. The coarse-grained phase space distribution function $ \bar f$ can therefore be seen as a theoretical N-body double. Another important property of $\bar f$ is that it can be obtained from $f_{\rm K}$ directly by coarse-graining in phase space, $\bar f = \exp\left[\tfrac{1}{2}\sigx^2\Delta_x + \tfrac{1}{2}\sigp^2\Delta_p\right] f_{\rm K}$,  without the need of  obtaining first $f$ and the Vlasov equation via ensemble averaging $f_{\rm K}$.

\subsection{Husimi-Vlasov equation}
\label{subsection:husimivlasovmap}
\subsubsection{Schr\"odinger Poisson system} \label{SPE}
The Schr\"odinger-Poisson system in a $\Lambda$CDM universe with scale factor $a$ is given by
\begin{subequations}
\label{schrPoissEqFRW}
\begin{align} 
i\hbar \del_t \psi &= - \frac{\hbar^2}{2a^2m} \Delta \psi + m V\psi \label{schrEqFRW} \,,\\
\Delta V&=\frac{4\pi G\,\rho_0}{a}\Big(|\psi|^2 -  1 \Big) \label{PoissEqFRW} \,,
\end{align}
\end{subequations}
see for instance \cite{WK93}.
Using the so-called Madelung representation for the wave function $\psi(\vx)=\sqrt{n(\vx)}\exp\left(i\phi(\vx)/\hbar\right)$ one can obtain fluid-like equations of motion for the normalized density\footnote{The volume average is $\langle n\rangle_{\rm vol}=1$.} $n$ and the velocity potential $\phi$ directly from the Schr\"odinger equation \cite{M27}. By separating real and imaginary parts one obtains the continuity equation \eqref{conti}, and an equation for $\phi$ which is similar to the Bernoulli equation \eqref{euler} but contains an extra  term proportional to $\hbar^2$, the so-called `quantum pressure'
\begin{subequations}
\label{MadelungFluid}
\begin{align} 
\del_t n &= -\frac{1}{m a^2}\vnabla\cdot (n \vnabla \phi)\label{MadelungConti}\,,\\
\del_t \phi &= -\frac{1}{2 a^2 m} \left(\vnabla\phi\right)^2 -mV +\frac{\hbar^2}{2a^2 m} \frac{\Delta\sqrt{n}}{\sqrt{n}} \label{MadelungEuler}\,, \\
\Delta V&=\frac{4\pi G\,\rho_0}{a}\Big(n -  1 \Big) \label{PoissEqFRW}  \,.
\end{align}
\end{subequations}
With the definition $\v{u}=\vnabla \phi/m$, the modified Bernoulli equation for $\phi$ is then equivalent to a modified Euler equation with the constraint $\vnabla \times \v{u} = 0$
\begin{subequations} \label{Madelungfluideq}
\begin{align}
\del_t n &= -\frac{1}{a^2}\vnabla_{\! \!  x}\cdot (n\v{u}) \,,\\
 \del_t \v{u} &= -\frac{1}{a^2}(\v{u}\cdot\vnabla)\v{u} -\vnabla V +\frac{\hbar^2}{2a^2 m^2} \vnabla \left( \frac{\Delta\sqrt{n}}{\sqrt{n}} \right) \label{EulerMadelung}\,.
\end{align}

At this stage we want to emphasize again that the Schr\"odinger equation is considered here as a mere tool to model CDM dynamics. Therefore the value of $\hbar$ has to be treated as a parameter which is not necessarily connected to the value of $\hbar$ in the context of ordinary quantum mechanics, but rather must be adjusted to computational feasibility and the physical problem at hand \cite{WK93}. Another important remark is in order. The Madelung respresentation Eqs.\,\eqref{MadelungFluid} is only equivalent to the Schr\"odinger system Eqs.\,\eqref{schrPoissEqFRW} as long as $n\neq 0$. We will see later that during shell-crossings interference in the wave-function $\psi$ will cause $n=0$ at isolated points in space and time. Once this happens the Madelung representation breaks down because $\phi$ develops infinite spatial gradients and phase jumps, leading to infinite time derivatives.  In App.\,\eqref{sec:lagrange} we investigate the Lagrangian formulation of the SPE, which suffers from the same problem. If one still prefers to stay in the fluid picture, one needs to solve instead for the momentum $\v{j}\equiv n \v{u}$, which is well behaved during these phase jumps and fulfills 
\begin{align}
\del_t n &= -\frac{1}{a^2}\vnabla\cdot \v{j} \,,\\
 \del_t \v{j} &= -\frac{1}{a^2}\nabla_i  \left(\frac{j_i \v{j}}{n}\right) -n \vnabla \left( V -\frac{ \hbar^2}{2a^2 m^2} \frac{\Delta\sqrt{n}}{\sqrt{n}} \right) \label{Eulerj}\,.
\end{align}
\end{subequations}
We will comment on the nature of phase jumps in Sec.\,\ref{timeevo}. The dynamics of $\psi$ in Eqs.\,\eqref{schrPoissEqFRW} is free from pathologies.

\subsubsection{Wigner quasi-probability distribution}
Originally introduced to study quantum corrections to classical statistical mechanics, the Wigner quasi-probability distribution \cite{W32} allows to link the Schr\"odinger wave function $\psi(\vx)$ to a function $f(\vx,\vp)$ in phase space
\begin{align}
\label{fWigner}
f_\W(\vx,\vp) =\int \frac{\vol{3}{\tilde x}}{(\pi\hbar)^3} \exp\left[2\frac{i}{\hbar}\vp\cdot\tilde \vx\right]\psi(\vx-\tilde \vx)\psi^*(\vx+\tilde \vx) \,, 
\end{align}
where $\psi^*$ denotes the complex conjugate of $\psi$. $f_\W$ is a quasi-probability distribution since it can become negative in general. For the dust-like initial conditions studied later see Fig.\,\ref{fig:wignerplot}, left.

\paragraph*{Wigner Vlasov equation}
The time evolution equation for $f_\W$ is obtained by using the Schr\"odinger equation \eqref{schrEqFRW} and performing an integration by parts twice which yields
\begin{eqnarray}
\partial_t f_\W  &&=  -\frac{\vp}{a^2m}\vnabla_{\! \!  x} f_\W +\frac{i}{\hbar} \int \frac{\vol{3}{\tilde x}}{(\pi\hbar)^3} \exp\left[2\frac{i}{\hbar}\vp\cdot\tilde \vx\right]\times \\
\notag &&\qquad\qquad \times\ m \left[V(\vx+\tilde\vx)-V(\vx-\tilde\vx)\right] \psi(\vx-\tilde\vx) \psi^*(\vx+\tilde\vx) \,.
\end{eqnarray}
In order to obtain a factorization of the form $V(\vx)\cdot f_\W$ one has to perform a Taylor expansion of $V(\vx-\tilde\vx)-V(\vx+\tilde\vx)$ around $\vx$ using $\alpha \in \mathbb N_0^3$ as a multi-index
\begin{align}
V(\vx+\tilde\vx)-V(\vx-\tilde\vx)&= \sum_{|\alpha|\geq 1} \frac{\del_x^{(\alpha )}V(\vx)}{\alpha !}\left[\tilde\vx^\alpha -(-\tilde\vx)^\alpha \right] \,.
\end{align}
Obviously the difference in parenthes vanishes if $|\alpha|$ is even and gives $2\tilde\vx^\alpha$ if $|\alpha|$ is odd. Therefore this term can be rewritten as derivative $-i\hbar \del_p^{(\alpha )} \exp\left[2i\vp\cdot \tilde\vx/ \hbar\right]$.
 Upon resummation one obtains the evolution equation for the Wigner function
\begin{subequations}
\label{WignerVlasov} 
\begin{align}
\partial_t f_\W &= -\frac{\vp}{a^2 m}\vnabla_{\! \!  x} f_\W + m V\frac{2}{\hbar}\sin\left(\frac{\hbar}{2}\overleftarrow{\vnabla}_{\! \!  x}\overrightarrow{\vnabla}_{\! \!  p}\right) f_\W \label{WignerVlasov1} \,,\\
&= \left[ \frac{\vp^2}{2a^2m}+m V\right]   \frac{2}{\hbar} \sin\left(\frac{\hbar}{2}( \overleftarrow{\vnabla}_{\! \!  x} \overrightarrow{\vnabla}_{\! \!  p}- \overleftarrow{\vnabla}_{\! \!  p} \overrightarrow{\vnabla}_{\! \!  x})\right) f_\W  \label{WignerVlasov2} \,,
\end{align}
\end{subequations}
which coincides with the result given in \cite{T89} for the special case where $a=1$. Note that on an FRW space $a(t)$ is the scale factor with $t$ cosmic time, $\bm x$ comoving, $\bm p$ is the conjugate momentum 1-form and $V$ fulfills Eq.\,\eqref{PoissEqFRW}.
\paragraph*{Relation to $f_\d$}
The similarity between the equations \eqref{MadelungFluid} obtained from a Schr\"odinger wave function when decomposing it into modulus and phase $\psi=\sqrt{n}\exp\left(i\phi/\hbar\right)$ and the fluid equations \eqref{fluid} can also be understood from the point of view of distribution functions. 
Transforming variables $\tilde x \rightarrow \hbar \tilde x$ and adopting the shorthand notation $g^\pm=g(\vx\pm\hbar\tilde\vx)$ the Wigner function can be rewritten in the following form 
\begin{align*}
f_\W(\vx,\vp) &=\int \frac{\vol{3}{\tilde x}}{\pi^3} \sqrt{n^+n^-}  \exp\left[i\left(2\vp\cdot\tilde \vx+\frac{\phi^- -\phi^+}{\hbar}\right)\right]\,,
\end{align*}
which allows to examine the formal limit $\hbar \rightarrow 0$. Taylor-expanding $n^\pm$ and $\phi^\pm$ to leading non-vanishing order in $\hbar$ and evaluating the integral gives \cite{WK93}
\begin{align}
f_\W(\vx,\vp) &\stackrel{\hbar \rightarrow 0}{=} n(\vx)\delta_{\rm D} \Big( \vp-\vnabla\phi(\vx) \Big) = f_\d (\vx,\vp) \,. 
\end{align}
\paragraph*{Correspondence to Vlasov equation} At leading order, the Wigner Vlasov equation \eqref{WignerVlasov} differs from the Vlasov equation \eqref{VlasovPoissonEq} only by a term proportional to $\hbar^2$
\begin{align*}
\partial_t \left(f_\W-f\right) \simeq \frac{\hbar^2}{24}\del_{x_i}\del_{x_j}\vnabla_{\! \!  x}V\del_{p_i}\del_{p_j}\vnabla_{\! \!  p} f_\W+\mathcal{O}(\hbar^4) \,.
\end{align*}
Therefore one might hope that they are in good agreement. However, as was shown exemplarily in \cite{T89}, the correspondence between the time-evolution of the Wigner distribution $f_\W$ and Vlasov distribution function $f$ is in general very poor by virtue of the violent oscillations of $f_\W$ on scales $\hbar$, related to the fact that $f_\W$ can become negative. In this context one has to bear in mind that the semiclassical limit $\hbar \rightarrow 0$ is not meaningful in the sense that it does not drive the solution towards a classical one in a continuous way. \\

\subsubsection{Coarse-grained Wigner distribution function}
The so-called Husimi-Q \cite{H40} representation can be understood as a smoothing of the Wigner quasi-probability distribution \eqref{fWigner} by a Gaussian filter of width $\sigx$ and $\sigp$ in $\vx$ and $\vp$ space, respectively
\begin{align}
\label{fcgWigner}
 \bar f_\W &= \exp\left(\frac{\sigx^2}{2}\Delta_x+\frac{\sigp^2}{2}\Delta_p\right) f_\W \,.
\end{align}
In contrast to the Wigner distribution itself the coarse-grained version is a positive-semidefinite function if the filter is of appropriate size $\sigx\sigp \geq\hbar/2$ for a semi-classical description, see \cite{C75}. Note that for the FRW case, the form of $ \bar f_\W$ remains unchanged provided $\bm x$ is comoving and $\bm p$ is the conjugate momentum 1-form.
\paragraph*{Husimi-Vlasov equation}
The corresponding Husimi-Vlasov equation for the coarse-grained $f_\W$ is then easily obtained by acting with the coarse-graining operators onto Eq.\,\eqref{WignerVlasov} employing again the product rule \eqref{productrule}
\begin{widetext}
\begin{subequations}
\label{cgWignerVlasov} 
\begin{align}
\partial_t  \bar f_\W &= -\frac{\vp}{a^2 m}\vnabla_{\! \!  x}  \bar f_\W -\frac{\sigp^2}{a^2 m}\vnabla_{\! \!  x}\vnabla_{\! \!  p}  \bar f_\W  + m \bar V\exp(\sigx^2\overleftarrow{\vnabla}_{\! \!  x}\overrightarrow{\vnabla}_{\! \!  x})\frac{2}{\hbar}\sin\left(\frac{\hbar}{2}\overleftarrow{\vnabla}_{\! \!  x}\overrightarrow{\vnabla}_{\! \!  p}\right)  \bar f_\W \label{cgWignerVlasov1} \,,\\
&= \exp \left(\frac{\sigx^2}{2}\Delta_x + \frac{\sigp^2}{2}\Delta_p \right) \left[ \frac{\vp^2}{2a^2m}+m V(\vx)\right]  \exp \left(\sigx^2 \overleftarrow{\vnabla}_{\! \!  x} \overrightarrow{\vnabla}_{\! \!  x}+ \sigp^2 \overleftarrow{\vnabla}_{\! \!  p} \overrightarrow{\vnabla}_{\! \!  p}\right)  \frac{2}{\hbar} \sin\left(\frac{\hbar}{2}( \overleftarrow{\vnabla}_{\! \!  x} \overrightarrow{\vnabla}_{\! \!  p}- \overleftarrow{\vnabla}_{\! \!  p} \overrightarrow{\vnabla}_{\! \!  x})\right)  \bar f_\W \,. \label{cgWignerVlasov2} 
\end{align}
\end{subequations}
\end{widetext}
This equation is the generalization of the result given in \cite{T89} allowing for cosmological backgrounds, arbitrary potentials and smoothing scales $\sigx, \sigp$. It is the resummation of the equation given up to second order in $\sigx$ in \cite{SRvB89}, which we obtained by explicit calculation performed analogously to the one presented for $f_\W$.

In \cite{WK93} the Husimi representation was used instead, in which the wave function is represented in a (over-complete) basis of Gaussian wave packets
\begin{subequations}\label{Husimi}
\begin{align} 
\psi_{\rm H}(\v{x},\v{p}) &= \int \vol{3}{y} K_{\rm H}(\v{x},\v{y},\v{p}) \psi(\v{y}) \,, \\
K_{\rm H}(\v{x},\v{y},\v{p}) &= \frac{\exp\left[-\frac{(\v{x}-\v{y})^2}{4 \sigx^2} - \frac{i}{\hbar} \v{p}\cdot \left(\v{y} -\frac{1}{2}\v{x}\right)\right] }{\left( 2\pi \hbar \right)^{3/2}  \left(2 \pi \sigx^2 \right)^{3/4}}\,,
\end{align}
\begin{center}
\begin{figure*}[t]
\hspace{-0.2cm}\includegraphics[width=0.8\textwidth]{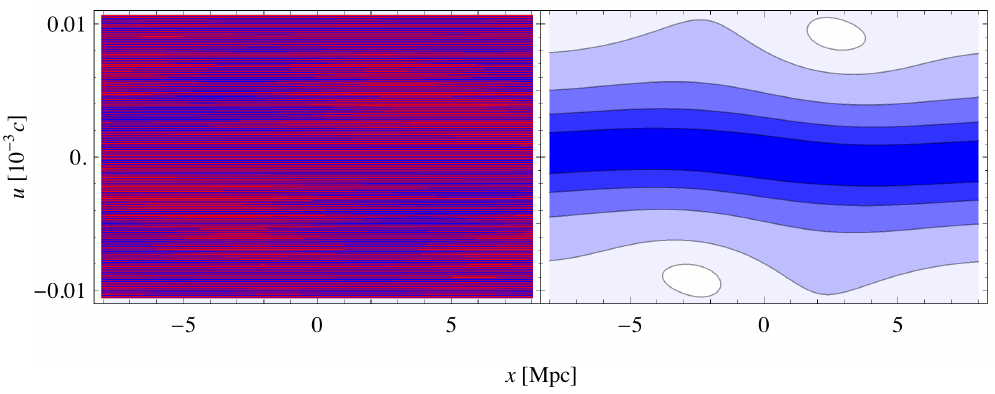}\\
\caption{Comparison between $f_\W$ and $\bar f_\W$ at the initial time $a_{\rm ini}=0.01$ using linear dust-like initial conditions $n=n_{\rm d}$ and $\phi= \phi_{\rm d}$. The left panel shows the strongly oscillating $f_\W$ (\textit{red} is negative, \textit{blue} is positive), the right panel is its smoothed version $\bar f_\W$ (with the same coloring as in Fig\,\ref{fig:phaseplot}).  We choose the minimal values of $\sigx$ and $\sigp$  such that $\bar f_\W \geq 0$, which turned out to be $\sigx \sigp = 0.03 \hbar $.}
\label{fig:wignerplot}
\end{figure*}
\end{center}
such that when going from $\psi$ to $\psi_{\rm H}$ no information is sacrificed. 
Defining the Husimi distribution function to be
\begin{equation}
f_{\rm H} =| \psi_{\rm H} |^2 \,,
\end{equation}
it is easy to check that it is a special case of the coarse-grained Wigner function, namely
\begin{equation}
f_\H= \bar f_\W \quad \text{if} \quad \sigx\sigp=\hbar/2 \,. \label{fHfW}
\end{equation}
\end{subequations}
This representation is very convenient numerically, because $f_\H$ is manifestly real and positive. Also the integration is much simpler to evaluate than for $f_\W$. The main advantage is that one does not need to sample the quite heavily oscillating $f_\W$  to construct $\bar f_\W$. Fig\,.\ref{fig:wignerplot} ({\it left}) provides an impression of $f_\W$ for cold initial conditions. We also know that $\sigx\sigp \geq\hbar/2$ ensures $\bar f_\W\geq 0$ \cite{C75}. Therefore the Husimi representation picks the smallest sufficient $\sigp$ for a positive phase space distribution given a $\sigx$ and $\hbar$. However we would like to point out that for cold dust-like initial conditions well within the linear regime we are free to choose even $\sigx\sigp < \hbar/2$ without encountering any trouble, compare Figs.\,\ref{fig:phaseplot} (1st panel) and \ref{fig:wignerplot} ({\it right}). It is also important to realize that the dynamics at early times well before shell-crossing is not affected by the seemingly poor phase space resolution, see Fig.\,\ref{fig:phaseplot} (1st panel). We can see this by inspecting the Madelung representation \eqref{MadelungFluid} of the Schr\"odinger equation from which it is clear that for smooth dust-like initial conditions the quantum potential 
 with
\begin{equation}
Q= - \frac{\hbar^2}{2a^2m^2} \frac{\Delta\sqrt n}{\sqrt n}\,,
\end{equation}
 will be subdominant for sufficiently small $\hbar/m$.

\paragraph*{Correspondence to coarse-grained Vlasov equation}
\label{coarseVlvsHus}
Comparing the coarse-grained Vlasov equation \eqref{cgVlasovEq} and the Husimi-Vlasov equation \eqref{cgWignerVlasov} we find that they are equal at first order in $\sigma^2_{\!\! x}$ and $\sigma^2_{\!\! p}$
\begin{eqnarray}
\partial_t \left( \bar f_\W- \bar f\right) \simeq \frac{\hbar^2}{24}\del_{x_i}\del_{x_j}\vnabla_{\! \!  x}\bar V\del_{p_i}\del_{p_j}\vnabla_{\! \!  p}  \bar f_\W+\mathcal{O}(\hbar^4,\hbar^2\sigx^2) \,. \qquad
\end{eqnarray}
The Husimi-Vlasov equation \eqref{cgWignerVlasov} is in good correspondence to the coarse-grained Vlasov equation \eqref{cgVlasovEq} if $\sigx\sigp \gtrsim \hbar/2$, which ensures the removal of the violent oscillations and therefore approximates the Vlasov equation well if $\sigx\ll x_{\text{typ}}$ and $\sigp\ll p_{\text{typ}}$. Hereby we compared the two distribution functions which are obtained with the same coarse-graining parameters $\sigx$ and $\sigp$ in phase space.
As described in \cite{T89},  the coarse-grained Wigner function $ \bar f_\W$ reveals a considerably better correspondence to the probability distribution function $f$ in classical mechanics than the Wigner function $f_\W$ does. 

\subsubsection{Appropriate choice of the smoothing scales}

If $x_{\rm typ}$ and $p_{\rm typ}$ are the (minimal) scales of interest we have to ensure that
\begin{align}
\sigx \ll x_{\text{typ}} \quad \mathrm{and} \quad \sigp \ll p_{\text{typ}} \,.
\end{align}
Furthermore in general the maximal achievable resolution in phase space is limited by the value of $\hbar$ such that $\sigx$ and $\sigp$ have to be chosen to fulfill
\begin{align}
\hbar/2 \lesssim \sigx\sigp \,,
\end{align}
see however Fig.\,\ref{fig:wignerplot} for an exception well before shell-crossing.
On a FRW background these bounds take the same form if distances are comoving and if $u_{\text{typ}} = p_{\text{typ}}/m$ is absolute value of the comoving (or canonical) momentum 1-form.  
For translating these bounds into requirements for numerical simulations, for example grid time resolution, we refer the reader to \cite{WK93}. 

 \section{Hierarchy of Moments}
\label{sec:Hierarchy}
In practice one is usually interested in following the evolution of the spatial distribution instead of describing the fully fledged phase space dynamics encoded in the Vlasov equation. For this purpose, the relevant information can be extracted by taking moments of the distribution function with respect to momentum.
\paragraph*{Generating functional} The moments $M^{(n)}$ of the phase space distribution function $f(\vx,\vp)$ can be obtained from the generating functional $G[\v{J}]$ by taking functional derivatives. In a similar way the cumulants can be determined from the moments. They provide a good way to understand the prominent dust-model which is the only known consistent truncation of the Vlasov hierarchy. The generating functional, moments and cumulants are given by
\begin{subequations}
\begin{align}
&G[\v{J}] = \int \vol{3}{p} \exp\left[i\vp\cdot\v{J}\right] f \,, \label{genfun}\\
&M^{(n)}_{i_1 \cdots i_n}:=\int \vol{3}{p} p_{i_1} \ldots p_{i_n} f = (-i)^n \left.\frac{\del^n G[\v{J}]}{\del J_{i_1} \ldots \del J_{i_n}} \right|_{\v{J}=0} \,, \label{moments}\\
&C^{(n)}_{i_1 \cdots i_n}:= (-i)^n \left.\frac{\del^n \ln G[\v{J}]}{\del J_{i_1} \ldots \del J_{i_n}} \right|_{\v{J}=0} \label{cumulants}\,.
\end{align}
\end{subequations}
\paragraph*{Vlasov hierarchy} The evolution equations for the moments $M^{(n)}$ of the phase space distribution $f$ can be determined from the Vlasov equation \eqref{VlasovEq} by multiplying it with $p_{i_1}\cdots p_{i_n}$ and performing an integration over momentum
\begin{align}
\label{VlasovHierarchy}
\partial_t M^{(n)}_{i_1 \cdots i_n} &=   - \frac{1}{a^2 m} \nabla_j M^{(n+1)}_{i_1 \cdots i_n j}  - m \nabla_{(i_1} V \cdot M^{(n-1)}_{i_2 \cdots i_n)} \,.
\end{align}
Indices enclosed in round brackets imply symmetrization according to $a_{(i}b_{j)}=a_ib_j+a_jb_i$.
It turns out that a coupled Vlasov hierarchy for the moments emerges which means that in order to determine the time-evolution of the $n$-th moment the $(n+1)$-th moment is required. This closure problem for the hierarchy becomes more transparent when looking at the dynamical equation for the $n$-th cumulant $C^{(n)}$. The time evolution can be determined from the generating functional \eqref{genfun} using the Vlasov equation \eqref{VlasovEq} and reads
\begin{align}
\notag \del_t C^{(n)} _{i_1\cdots i_n} &= -\frac{1}{a^2m} \Bigg\{ \nabla_j C^{(n+1)}_{i_1\cdots i_n j} +\sum_{S\in \mathcal P(\{i_1,\cdots,i_n\})} C^{(n+1-|S|)}_{l\notin S,j} \cdot \nabla_j C^{(|S|)}_{k\in S} \Bigg\}\\
&\quad - \delta_{n1} \cdot m \nabla_{i_1}V \,,
\end{align}
where $S$ runs through the power set $\mathcal P$ of indices $\{i_1,\cdots,i_n\}$ and the Kronecker $\delta_{n1}$ in last term ensures that the potential contributes only to the equation for the first cumulant $C^{(1)}$ describing velocity. From this equation it becomes clear that one can set $C^{(n\geq 2)} \equiv 0 $ in a consistent manner since each summand in the evolution equation of $C^{(2)}$ contains a factor of $C^{(n\geq 2)}$. In contrast, the time evolution of $C^{(3)}$ depends also on summands containing solely $C^{(2)}$ such that it cannot be trivially fulfilled when setting $C^{(n\geq 3)} \equiv 0$. A similar reasoning applies to all higher cumulants $C^{(n\geq 3)}$ and demonstrates that there is no consistent truncation of the hierarchy of cumulants apart from the one at second order. These arguments are seconded by numerical evidence indicating that as soon as velocity dispersion encoded in $C^{(2)}$ becomes relevant, even higher cumulants are sourced dynamically, see \cite{PS09}.

\paragraph*{Strategies for closing the hierarchy}
In principle it would be desirable to adopt an ansatz for $f$ as general as possible. However, in this case it is difficult to find a closed form expression for the moments since one cannot perform the integration over momentum space. Therefore we have to resort to a special ansatz for the $p$-dependence of $f$ which allows to compute moments up to arbitrary order analytically. In the following we will compare three different ans\"atze for the distribution function $f$: the dust model $f_\d$, the Wigner function $f_\W$ as well as the Husimi distribution function $ \bar f_\W$. 

\subsection{Hierarchy of moments of $f_\d$}
 The generating functional for the dust model where $f_\d$ was inserted according to \eqref{fdust} is given by
\begin{align}
&G_\d[\v{J}] =n_\d \exp\left[i\vnabla \phi_\d\cdot\v{J}\right] \,.
\end{align}
The moments $M_\d^{(n)}$ and cumulants $C_\d^{(n)}$ are then given by
\begin{subequations}
\begin{align}
M_\d^{(0)} &= n_\d\,, \qquad {M_\d}^{(1)}_i=n_\d \phi_{\d,i}  \,, \quad { M_\d}^{(n \geq 2)}_{i_1\cdots i_n} = n_\d \phi_{\d,i_1} \cdots \phi_{\d,i_n} \,,\\
C_\d^{(0)} &= \ln n_\d\,, \quad \ {C_\d}^{(1)}_i=\phi_{\d,i}  \,, \quad \quad \  {C_\d}^{(n \geq 2)}_{i_1\cdots i_n} = 0\,.
\end{align}
\end{subequations}
Since the exponent of the generating functional is manifestly linear in $\v{J}$, all cumulants of order higher than one vanish identically. This means that the dust model does not include effects like velocity dispersion, which is encoded in the second cumulant $C^{(2)}$, or vorticity since the velocity is determined from a potential $\phi$. Therefore for the dust ansatz $f_\d$, the Vlasov equation is equivalent to its first two equations of the hierarchy of moments, the pressureless fluid system \eqref{fluid} consisting of the continuity and Euler equation. The first two moments of the Vlasov hierarchy \eqref{VlasovHierarchy} are
\begin{subequations}
\label{Fluid}
\begin{align}
\del_t n_\d &=   -\frac{1}{a^2m}\nabla_k( n_\d \phi_{\d,k}) \,, \label{Conti}\\
\del_t (n_\d\phi_{\d,i}) &= -\frac{1}{a^2 m} \nabla_j \left[n_\d\phi_{\d,i}\phi_{\d,j} \right]- m n_\d \nabla_i  V_\d   \,. \label{Euler}
\end{align}
\end{subequations}
If $n_\d$ and $\phi_\d$ fulfill these equations then all evolution equations of the higher moments are automatically satisfied, for example Eqs.\,\eqref{Fluid} imply that
\begin{align} \label{consistency}
\del_t (n_\d\phi_{\d,i}\phi_{\d,j} )&= -\frac{1}{a^2  m}\nabla_k \left(n_\d\phi_{\d,i}\phi_{\d,j} \phi_{\d,k}\right) -  m  n_\d\nabla_{(i}V_\d \cdot \nabla_{j)}\phi_\d \,.
\end{align}

\subsection{Hierarchy of moments of $f_\W$}
For simplicity we first consider the Wigner distribution function $f_\W$ as a model for a general distribution function $f$ fulfilling the Vlasov equation. This case will serve as pedagogical demonstration how the closure of the hierarchy can be achieved by choosing a special ansatz for the distribution function.
The generating functional can be computed by plugging the expression for $f_\W$ in terms of $\psi = \sqrt n \exp\left(i\phi/\hbar\right)$ in \eqref{genfun} and simplified by adopting again the shorthand notation $g^\pm(\vx'):= g\left(\vx'\pm\frac{\hbar}{2}\v{J}\right)$
\begin{align}
G[\v{J}] &= \sqrt{n^+n^-}\ \exp\left[\tfrac{i}{\hbar}(\phi^+-\phi^-)\right] \,.
\end{align}
From this expression the calculation for the moments $M^{(n)}$ is straightforward and yields
\begin{subequations} \label{momentsw}
\begin{align}
M^{(0)} &= n\,, \qquad M^{(1)}_i=n \phi_{,i}  \,. \label{01momentw} 
\end{align}
As expected, even all higher moments $M^{(n\geq 2)}$ of $f_\W$ are given in terms of the two scalar degrees of freedom $n$ and $\phi$ introduced as modulus and phase of the wave function $\psi$, respectively
\begin{align}
M^{(2)}_{ij} &=     n \left[ \phi_{,i} \phi_{,j} + \frac{\hbar^2}{4}  \left(\frac{n_{,i}n_{,j}}{n^2} - \frac{n_{,ij}}{n}\right)\right]\,, \label{2momentw} \\
M^{(3)}_{ijk}&= n \left[\phi_{,i}\phi_{,j}\phi_{,k}  + \frac{\hbar^2}{4} \left( \stackrel{+\text{cyc. perm.} }{\left(\frac{n_{,i}n_{,j}}{n^2} - \frac{n_{,ij}}{n}\right)\phi_{,k}}  -\phi_{,ijk} \right)\right]   \label{3momentw} \,,\\
 \sigma_{ij}&:=  \frac{\hbar^2}{4}  \left(\frac{n_{,i}n_{,j}}{n^2} - \frac{n_{,ij}}{n}\right) =C^{(2)}_{ij}\,, \quad C^{(3)}_{ijk}= - \frac{\hbar^2}{4} \phi_{,ijk}  \,.
\end{align}
\end{subequations}
To those terms which are marked by `+ cyc. perm.' cyclic permutations of the indices have to be added. As we will explain in the following, this special form of the higher moments and cumulants amounts to having closed the infinite Wigner-Vlasov hierarchy for the moments of $f_\W$ without truncating it. To demonstrate this we take moments of the Wigner-Vlasov equation \eqref{WignerVlasov} where we consider corrections to the Vlasov equation up to arbitrary order in $\hbar^2$. The $\hbar$-terms constitute correction terms to the Vlasov hierarchy \eqref{VlasovHierarchy} which become relevant for $M^{(n\geq 3)}$ but do not contribute to $M^{(n\leq 2)}$ since they have at least three derivatives with respect to momentum which cancel all lower moments than the third. Therefore the first three evolution equations are completely analogous to the ones obtained for the dust model. By plugging in the expression for $M^{(2)}$ we obtain a closed system of differential equations for $n$ and $\phi_{,i}$. 
\begin{subequations}
\label{FluidW}
\begin{align}
&\del_t n =   -\frac{1}{a^2m}\nabla_k( n \phi_{,k}) \,, \label{ContiW}\\
&\del_t (n\phi_{,i}) = -\frac{1}{a^2 m} \nabla_j \left[n\phi_{,i}\phi_{,j} + \frac{\hbar^2}{4} \left(\frac{n_{,i}n_{,j}}{n}-n_{,ij}\right)\right]  - n m\nabla_i V   \,.\label{EulerW}
\end{align}
\end{subequations}
We see that Eqs.\,\eqref{FluidW} determining time evolution of the first two moments of $f_\W$ are identical to the fluid-like equations obtained directly from the Schr\"odinger equation \eqref{Madelungfluideq}. This can be verified easily by plugging \eqref{ContiW} into \eqref{EulerW} and using that the difference in the 'quantum velocity dispersion' term arising from \eqref{2momentcgw} and \eqref{MadelungEuler} is only apparent since 
\begin{align} \label{qpressure}
\frac{\hbar^2}{4} \nabla_j\left(\frac{n_{,i}n_{,j}}{n}-n_{,ij}\right) = -\frac{\hbar^2}{2} n \nabla_i \left(\frac{\Delta \sqrt n}{\sqrt n}\right) \,.
\end{align}
Note that a proper pressure term in the Euler equation would have the form $\v{\nabla} p$ with some $p$ and not the form $ n \v{\nabla} Q$.  Rather the left hand side of \eqref{EulerW} suggests that this term constitutes a 'quantum velocity dispersion', since it is of the form $\nabla_i (n\sigma_{i j})$. Equivalently, one can interpret the $\hbar$ term as a correction to the Newtonian potential $V\rightarrow V +Q$.

The evolution equation for the second moment $M^{(2)}$ involves the third moment $M^{(3)}$ and is given by
\begin{align} \label{consistencyW}
\del_t M^{(2)}_{ij}&= -\frac{1}{a^2  m}\nabla_k M^{(3)}_{ijk} - n m \nabla_{(i} V \cdot \nabla_{j)}\phi \,.
\end{align}
For the Wigner function $f_\W$ all moments $M^{(n)}$ can be expressed entirely in terms of the density $n$ and conjugate velocity $\vnabla \phi$. Hence, this ansatz closes the hierarchy since Eq.\,\eqref{consistencyW} is automatically fulfilled when $M^{(2)}$ and $M^{(3)}$, taken from \eqref{2momentw} and \eqref{3momentw} respectively, are expressed in terms of $n$ and $\phi$ which fulfill the corresponding fluid equations \eqref{FluidW}. The same is true for all higher moments.

\subsection{Hierarchy of moments of $ \bar f_\W$} \label{HusHierarchy}
\subsubsection{Moments up to third order}
We want to resort to a special ansatz for the $p$-dependence of $f$ which allows to compute moments up to arbitrary order analytically. The coarse-grained Wigner distribution function $ \bar f_\W$ provides us with such an ansatz. Furthermore it is well-suited to model a general distribution function $f$ fulfilling the Vlasov equation as was demonstrated in \cite{WK93}. By plugging in the expression for $ \bar f_\W$ in terms of $\psi = \sqrt n \exp\left(i\phi/\hbar\right)$ we can rewrite the generating functional to get
\begin{align}
\label{genfuncgW}
\bar G[\v{J}] &= \exp\left[\tfrac{\sigx^2}{2}\Delta-\tfrac{\sigp^2}{2} \v{J}^2\right] \sqrt{n^+n^-}\ \exp\left[\tfrac{i}{\hbar}(\phi^+-\phi^-)\right] \,.
\end{align}
From this expression the calculation for the moments $\bar M^{(n)}$ is straightforward and yields
\begin{subequations} \label{momentscgw}
\begin{align}
\bar M^{(0)} &=: \bar n=  \exp\left(\frac{\sigx^2}{2} \Delta\right) n \label{0momentcgw} \,, \\
\bar M^{(1)}_i&=:m \bar n \bar u_i =   \exp\left(\frac{\sigx^2}{2} \Delta\right) \left(n \phi_{,i}\right) \label{1momentcgw} \,.
\end{align}
The corresponding mass weighted velocity $\bar{\v{u}}$ is obtained by smoothing the momentum field and then dividing by the smoothed density field. This is precisely the definition commonly used in the effective field theory of large-scale structure, compare \cite{M13, B10}. From a physical point of view $\bar{\v{u}}$ describes the center-of-mass velocity of the collection of particles inside a coarsening cell of diameter $\sigx$ around $\v{x}$.\\
As expected, even all higher moments $\bar M^{(n\geq 2)}$ of $ \bar f_\W$ are given in terms of the two scalar degrees of freedom $n$ and $\phi$ introduced as modulus and phase of the wave function $\psi$, respectively
\begin{align}
\bar M^{(2)}_{ij} &=    \exp\left(\frac{\sigx^2}{2} \Delta\right) \left\{ n \left[ \phi_{,i} \phi_{,j} +\sigp^2\delta_{ij} + \sigma_{ij}\right]\right\}  \label{2momentcgw} \,,\\
\bar M^{(3)}_{ijk}&=  \exp\left(\frac{\sigx^2}{2} \Delta\right) \left\{ n\left[ \phi_{,i}\phi_{,j}\phi_{,k} + \stackrel{+\text{cyc. perm.}}{\left(\sigp^2\delta_{ij}+ \sigma_{ij}\right)\phi_{,k}} - \frac{\hbar^2}{4} \phi_{,ijk} \right] \right\}   \,. 
 \label{3momentcgw}
\end{align}
The corresponding cumulants can be calculate from the previous results using
\begin{align}
\bar{C}^{(2)}_{ij} &= \frac{\bar M^{(2)}_{ij}}{\bar M^{(0)}} - \frac{\bar M^{(1)}_i\bar M^{(1)}_j}{[\bar M^{(0)}]^2} \\
&= \sigp^2 \delta_{ij} + \frac{\overline{n\sigma_{ij}}}{\bar n} + \frac{\overline{n\phi_{,i}\phi_{,j}}}{\bar n} - \frac{\overline{n\phi_{,i}}\ \overline{n\phi_{,j}}}{\bar n^2} \label{barC2} \,,\\
\bar{C}^{(3)}_{ijk} &= \frac{\bar M^{(3)}_{ijk}}{\bar M^{(0)}} - \stackrel{+ \text{cyc. perm.}}{\frac{\bar M^{(2)}_{ij}\bar M^{(1)}_k}{[\bar M^{(0)}]^2}}+2 \frac{\bar M^{(1)}_i\bar M^{(1)}_j\bar M^{(1)}_k}{[\bar M^{(0)}]^3} \\
&= \frac{\bar M^{(3)}_{ijk}}{\bar M^{(0)}} - \stackrel{+ \text{cyc. perm.}}{\bar C^{(2)}_{ij}\bar C^{(1)}_k}- \bar C^{(1)}_i\bar C^{(1)}_j\bar C^{(1)}_k \,.
\end{align}
\end{subequations}
As we will explain in the following, this allows to close the infinite hierarchy for the moments of $ \bar f_\W$ arising from the Husimi-Vlasov Eq.\,\eqref{cgWignerVlasov} without setting any of the cumulants to zero. Instead, all higher moments are determined self-consistently from the lowest two, which are dynamical and represent the coarse-grained density $\bar n$ and velocity $\bar{\v{u}}$, respectively. This distinguishes our formalism fundamentally from phenomenological models which attempt to close the hierarchy by postulating an ansatz for the second cumulant, called stress tensor $n\sigma_{ij}$, but simultaneously setting all higher cumulants to zero. For example, the ansatz for the velocity dispersion of a cosmological imperfect fluid is given by $n\sigma_{ij}= p\delta_{ij} +\eta(\nabla_iu_j \nabla_ju_i -\frac{2}{3}\delta_{ij} \nabla \cdot u )+\zeta \delta_{ij} \nabla\cdot u$ where $p$ denotes the pressure and $\eta$ and $\zeta$ are shear and bulk viscosity coefficients respectively. The underlying approximation $\sigma_{ij} \approx 0$ is valid during the first stages of gravitational instability when structures are well described by a single coherent flow (single-stream). However, as soon as multiple streams become relevant after shell-crossing, velocity dispersion and vorticity are generated dynamically and at once all higher moments become relevant too \cite{PS09}. Thus, the hierarchy of cumulants of CDM dynamics cannot be truncated after shell-crossing has occurred. \\

In the subsequent calculation it will be necessary to reexpress all higher moments entirely in terms of  $\bar M^{(0)}\propto\bar n $ and $\bar M^{(1)}\propto \bar n \bar u_i$. For this purpose we introduce the $D$-symbol which allows us to express the coarse-graining of any product or quotient entirely in terms of its coarse-grained constituents, for example 
\begin{align}
\label{Dop}
\exp\left[\tfrac{1}{2}\sigma^2 \Delta\right]\left(n\phi_{,i}\phi_{,j}\right)=\exp\left[\tfrac{1}{2} \sigma^2  (\Delta-D)\right]\left( \frac{(\bar n \bar u_i)(\bar n \bar u_j)}{\bar n}\right) \,.
\end{align}
\subsubsection{Properties of the $D$-symbol}

$D$ fulfills the Leibniz product rule of a first derivative operator when acting on compositions of $$A, B, C \in \{\bar n, \bar n \bar u_i\} $$ or derivatives thereof, but when acting on a single function it is the Laplacian. 
\begin{subequations}
\label{Doperator}
\begin{align}
D(A) &= \Delta A \ \,,\  D(g(A)) = \partial_{A}g(A) DA =  \partial_{A} g(A) \Delta A \,,\\
D(AB) &= (DA) B+ A (DB) = (\Delta A) B+ A (\Delta B) \,.
\end{align}
\end{subequations}
Applying the definition of the $D$-symbol one can derive the following expressions for the evaluation of $D^n$ 
\begin{subequations}
\begin{align}
D^n \left( \frac{AB}{C} \right)&=  \sum\limits_{k=0}^{n} \binom{n}{k} \sum\limits_{l=0}^{n-k} \binom{n-k}{l} \Delta^l A \cdot  \Delta^{n-k-l}B \cdot D^k \left(\frac{1}{C} \right) \,,\\ 
D^k \left(\frac{1}{C}\right) &= \sum\limits_{r=0}^k \frac{(-1)^r r!}{C^{r+1}}B_{k,r}\left(\Delta C, \Delta^2 C,...,\Delta^{k-r+1} C\right)\,,
\end{align}
where $B_{k,r}$ are the Bell polynomials. Furthermore we have that
\begin{align}
\frac{1}{\exp\left(\sigx^2\Delta\right)C} &= \exp\left(\sigx^2D\right) \left(\frac{1}{C}\right) \,.
\end{align}

\end{subequations}

\subsubsection{Evolution equations for the moments of $ \bar f_\W$}

We take moments of the Husimi-Vlasov equation where we consider corrections to the Vlasov equation up to arbitrary order in $\sigx^2$, $\sigp^2$ and $\hbar^2$.  Eq.\,\eqref{cgWignerVlasov} can be employed to obtain evolution equations for the first two moments $\bar n = \bar M^{(0)}$ and $\bar u_i=\bar M^{(1)}_i/(m\bar n) $ which correspond to density and mass-weighted velocity, respectively. The velocity $\bar u_i$ which follows from a coarse-grained distribution function $\bar f$ is automatically a mass-weighted one computed according to $m \bar u_i = \overline{n\phi_{,i}}/\bar n$ and does not coincide with the volume-weighted velocity $\bar \phi_{,i}$. In particular, the volume-weighted velocity is automatically curl-free whereas the mass-weighted velocity will have vorticity in general.\\
By plugging in the expression for $\bar M^{(2)}$ and rewriting it according to \eqref{Dop} we obtain a closed system of differential equations for $\bar n$ and $\bar u_i$
\begin{subequations}
\label{FluidcgW}
\begin{align}
\del_t \bar n &=   -\frac{1}{a^2}\v{\nabla}\cdot(\bar n \bar{\v{u}}) \,, \label{ConticgW}\\
\del_t (\bar n \bar u_i) &= -\frac{1}{a^2 m^2} \nabla_j \bar  M^{(2)}_{ij} - \nabla_i \bar V \exp\left(\sigx^2\overleftarrow{\vnabla}_{\! \!  x}\overrightarrow{\vnabla}_{\! \!  x}\right) \bar n + \frac{\sigp^2}{a^2 m^2} \nabla_i \bar n \,,\notag\\
&=\exp\left(\frac{\sigx^2}{2} (\Delta-D)\right)\notag\Bigg\{-\frac{1}{a^2 m^2}\nabla_j \Bigg[ \frac{(\bar n \bar u_i) (\bar n\bar u_j)}{\bar n} +\\
&\qquad \quad+ \frac{\hbar^2}{4} \left(\frac{\bar n_{,i} \bar n_{,j}}{\bar n}-\bar{n}_{,ij}\right) \Bigg] - \bar n\ \nabla_i \bar V  \Bigg\}\,, \label{EulercgW}\\
\Delta \bar V&=\frac{4\pi G\,\rho_0}{a}\Big(\bar n -  1 \Big) \label{PoissEqcgW}  \,.
%
\end{align}
These equations are supplemented by the constraint that there exists a scalar function $ \bar \phi$ such that
 \begin{equation}
 m\, \bar n\, \bar{\v{u}} = \bar n\exp\left(\sigx^2\overleftarrow{\vnabla}_{\! \!  x}\overrightarrow{\vnabla}_{\! \!  x}\right) \vnabla \bar \phi\,. \label{Husimiconstr}
  \end{equation}
\end{subequations}
The last constraint equation is the analogue of the curl-free constraint Eq.\,\eqref{dustconstr}. It enforces a very particular non-zero vorticity for $\bar{\v{u}}$.
\label{3rdmoment}
The evolution equation for the second moment $\bar M^{(2)}$ involves the third moment $\bar M^{(3)}$ and is given by
\begin{align} \label{consistencycgW}
\del_t \bar M^{(2)}_{ij}&= -\frac{1}{a^2  m}\nabla_k \bar M^{(3)}_{ijk} - m \nabla_{(i}\bar V  \exp\left(\sigx^2\overleftarrow{\vnabla}_{\! \!  x}\overrightarrow{\vnabla}_{\! \!  x}\right) (\bar n \bar u_{j)})\\
\notag &\quad + \frac{\sigp^2}{a^2} (\bar n \bar u_{(i})_{,j)} \,.
\end{align}
For the coarse-grained Wigner distribution function $ \bar f_\W$ all moments $\bar M^{(n)}$ can be expressed entirely in terms of the density $\bar n$ and velocity $\bar{\v{u}}$. This ansatz closes the $\bar f_\W$ hierarchy since all higher moment equations are automatically fulfilled when $\bar M^{(n)}$ is calculated from \eqref{genfuncgW}, expressed in terms of $\bar n$ and $\bar{\v{u}}$ which are to be determined from the coarse-grained fluid equations \eqref{FluidcgW}. In appendix \ref{closeHierarchy} we show  by explicit computation that Eq.\,\eqref{consistency} is automatically satisfied when $\bar M^{(2)}$ and $\bar M^{(3)}$ are taken from \eqref{2momentcgw} and \eqref{3momentcgw} respectively.\\
Alternatively and for practical applications, instead of solving the coarse-grained fluid equations \eqref{FluidcgW} for $\bar n$ and $\bar {\v{u}}$ one can simply solve the SPE \eqref{schrPoissEqFRW} for $n$ and $\phi$ and construct the cumulants of interest according to \eqref{momentscgw}. Both procedures automatically and self-consistently include multi-streaming effects. Note that Eqs.\,\eqref{FluidcgW} are naturally written in terms of the macroscopic momentum $\bar{\v{j}} \equiv \bar n \bar{\v{u}}$, which is just the coarse-grained quantum momentum and therefore free from phase jump pathologies, see Sec.\,\ref{SPE}.
\vspace{-0.8cm}
\subsection{Comparison between the models}
If we compare the fluid equations obtained via the Husimi approach Eqs.\,\eqref{FluidcgW} with the one obtained directly from the Madelung representation Eqs.\,\eqref{MadelungFluid} of the underlying Schr\"odinger-Vlasov system we see that our special ansatz for the distribution function $f= \bar f_\W$ amounts to considering a spatially coarse-grained Schr\"odinger-Vlasov system.  However, we have to bear in mind that this is not equivalent with a direct coarse-graining of $n$ and $\phi_{,i}$ since the mass-weighted velocity is $m \bar u_i = \overline{n\phi_{,i}}/\bar n$ is not the same as the volume-weighted velocity $\bar \phi_{,i}$.
It is nontrivial that although $ \bar f_\W$ is coarse-grained with respect to space and momentum, the Schr\"odinger equation \eqref{MadelungFluid} and the first moment equations \eqref{FluidcgW} of $ \bar f_\W$ are related only by spatial coarse-graining. 
Note however that for instance the velocity dispersion $ \bar C^{(2)}_{ij}$ does depend on $\sigp$ as well as on $\sigx$ and $\hbar$, see Eq.\,\eqref{2momentcgw}.

One the one hand, by neglecting the $\hbar$-corrections which constitute a `quantum velocity dispersion' term in the Euler-type equation in \eqref{EulercgW} we obtain the same evolution equations for the coarse-grained fields $\bar n$ and $\bar{\v{u}}$ as given in \cite{D00,BD05}. Their approach started from a microscopic system of $N$ particles, which was spatially coarse-grained to obtain a set of hydrodynamic equations for the macroscopic fluid variables $\bar n$ and $\bar{\v{u}}$. This was done by expanding the smoothing operator $\exp\left[\frac{1}{2}\sigx^2\Delta\right]$ up to first order in the so-called large-scale expansion. Interestingly, these closed-form equations can be derived from our formalism based on the Schr\"odinger equation when setting $\hbar \rightarrow 0$ in \eqref{EulercgW}
\begin{align}
\label{Eulercg}
&\del_t (\bar n \bar u_i) = \exp\left(\frac{\sigx^2}{2} (\Delta-D)\right) \Bigg\{ -\frac{1}{a^2 m^2}\nabla_j \left[ \frac{(\bar n \bar u_i) (\bar n\bar u_j)}{\bar n}  \right] - \bar n\ \nabla_i \bar V  \Bigg\}\,. \notag
\end{align}
In this sense we provide a formal resummation in the large-scale parameter of \cite{D00}. Furthermore we can clearly see that one would have arrived exactly at same equation by spatially coarse-graining a dust fluid \eqref{Fluid}. However, this identification is only meaningful as long as no shell-crossing has occurred in the microscopic dust fluid as otherwise the filtering cannot be inverted. This explains the apparent contradiction between the fact that the dust model breaks down at shell-crossing although, according to \cite{D00}, the macroscopic system shows adhesive behavior. Obviously, the exact dust solution extended after shell-crossing, see red dashed line in Fig.\,\ref{fig:phaseplot}, does not exhibit adhesive behavior and coarse-graining cannot change this. This exemplifies that it is no longer possible to obtain the macroscopic quantities as the coarse-grained solution to the microscopic dust equations \eqref{Fluid}.

On the other hand, numerical examples show that the $\hbar$-term in the ScM regularizes shell-crossing caustics already on the microscopic level, see \cite{T89, SC02} and the next section. This allows to derive \eqref{FluidcgW} from the SPE \eqref{schrPoissEqFRW} and shows that in order to obtain a solution to the macroscopic system \eqref{FluidcgW} one can simply coarse-grain the solution to the microscopic system. Therefore the Schr\"odinger method may be viewed as improved dust model with  built-in infinity regularization (quantum potential proportional to $\hbar^2$ in \eqref{Eulerj})  as well as built-in eraser of regularization artefacts (spatial coarse-graining with $\sigx$ in \eqref{FluidcgW}). 

 Nearly cold initial conditions can be implemented by choosing 
\begin{equation}\label{inipsi}
\psi_{\rm{ini}}(x)= \sqrt{n_{\rm d}(a_{\rm{ini}},x)} \exp\left[ i \phi_{\rm d}(a_{\rm{ini}},x)/\hbar \right]\,,
\end{equation}
at some early time where shell-crossings have not occurred yet, where $n_{\rm d}$ and $\phi_{\rm d}$ denote solutions to the dust system \eqref{dustEqs}.
Although we have our focus on cold dark matter, let us remark that ScM also opens up the possibility to study warm initial conditions.

\vspace{-0.7cm}
\section{Numerical example} \label{sec:numerics}
We study the standard toy example of sine wave collapse, whose exact solution up to shell-crossing is given by the (in this case exact) Zel'dovich approximation \cite{Z70} and therefore has a long tradition in testing techniques of LSS calculations \cite{KS83}. Of particular relevance to our work is \cite{SC02}  were the collapse of a wave function fulfilling the Schr\"odinger Poisson equation and modifications to it were studied and compared to the exact Zel'dovich solution.
\vspace{-0.6cm}
\subsection{Initial conditions}
As reviewed in App.\,\ref{sec:lagrange}, the Zel'dovich approximation in the 1D (or plane parallel or pancake) collapse is the exact solution to the hydrodynamic  Eqs.\,\eqref{Fluideq}. We choose as initial linear density contrast
\begin{subequations}\label{delini}
\begin{equation} 
\delta_{\rm {lin}}(a,q) = D(a) \cos\left(\frac{\pi q}{L}\right) \,,
\end{equation}
which guarantees collapse at $a=1$, because according to Eq.\,\eqref{Zeldosol} the nonlinear density for dust is given by 
\begin{equation}
n_{\rm d}(a,q)= [1-\delta_{\rm lin}(a,q) ]^{-1}
\end{equation}
 choosing $D(1)=1$. The displacement field $\varPsi$ describes the trajectories $x(q)=q+\varPsi(a,q)$ of fluid elements and is given by
\begin{equation}
\varPsi_{\rm d}(a,q) = - D(a)\frac{L}{\pi} \, \sin\left(\frac{\pi q}{L} \right) \,,
\end{equation}
which can be used to express the velocity
\begin{equation}
\partial_x \phi_{\rm d} = u_{\rm d}(q)=a^3 H(a) \partial_a \varPsi_{\rm d}(a,q)
\end{equation}
\end{subequations}
 and density $n_{\rm d}$ in terms of $x$. We choose an Einstein-de Sitter universe, $H^2 = 8 \pi G/3\, \rho_0 a^{-3}$ with $H(a\!=\!1)= 70\,\mathrm{km}\,\mathrm{s}^{-1}\mathrm{Mpc}^{-1}$ and we pick $L=10\,\mathrm{Mpc}$.

We start to solve the Schr\"odinger equation at $a_{\rm{ini}}=0.01$ and choose as initial wave function Eq.\,\eqref{inipsi} with periodic boundary conditions such that $-L< x <L$.
We verified that during the linear stage of collapse, the phase $\phi$ and amplitude $n$ of the wave function, agree with their dust analogues $\phi_{\rm d}$ and $n_{\rm d}$  if $\tilde\hbar \equiv \hbar/m \lesssim 10^{-4}\,\mathrm{Mpc}\,c$, where $c$ is the speed of light. This agrees with findings of \cite{SC02}. In the remaining section we will mostly show results for $\tilde\hbar= 2\times10^{-5}\,\mathrm{Mpc}\,c$ and $\sigx=0.1\,\mathrm{Mpc}$. Only for the study of relaxation ($a=30.0$ in the following plots) as well as the Bohmian trajectories -- the integral lines of $\partial_x \phi$ -- in App.\,\ref{sec:lagrange} we choose the larger value $\tilde{\hbar} = 10^{-4}\,\mathrm{Mpc}$ and $\sigx=0.2\,\mathrm{Mpc}$.
Note that the mass $m$ can be absorbed in $\phi$ and $\phi_{\rm d}$, whereby $m$ disappears from the Schr\"odinger and fluid equations, respectively. The Wigner and coarse-grained Wigner functions are depicted in Fig.\,\ref{fig:wignerplot}.

It turns out that in single-streaming regions one can choose $\sigx \sigp \ll \hbar$ while still ensuring $\bar f_\W \geq 0$, see Fig.\,\ref{fig:wignerplot}. Comparing to the top panel of Fig.\,\ref{fig:phaseplot}, it becomes clear that  $\bar f_\W$  can achieve a much higher resolution than $f_{\rm H}$ in $u$-direction. It exemplifies that the initial conditions are well modeled by the SPE and that the large width of $f_{\rm H}$ in the initial conditions shown in Fig.\,\ref{fig:phaseplot} does not imply that the dynamics is poorly resolved. In contrast, it only means that if we want to use the more convenient $f_{\rm H}$ we sacrifice available information once we calculate moments and cumulants. Another possibility to circumvent the oscillatory behaviour of the Wigner function is to use a mixed state corresponding to $N$ gravitating wave functions rather than a single one. This was the method of choice in \cite{SBRB13}. It turns out that if $N$ is large enough, the Wigner function becomes well behaved even without any smoothing. Since our goal is to develop analytical tools on the basis of the ScM, is seems to be more prospective to consider a single wave function and adopt the Husimi representation.

\subsection{\boldmath Time evolution of  $\psi$, $f_{\rm H}$ and moments}
\label{timeevo}

\begin{center}
\begin{figure}[t]
\includegraphics[width=0.41\textwidth]{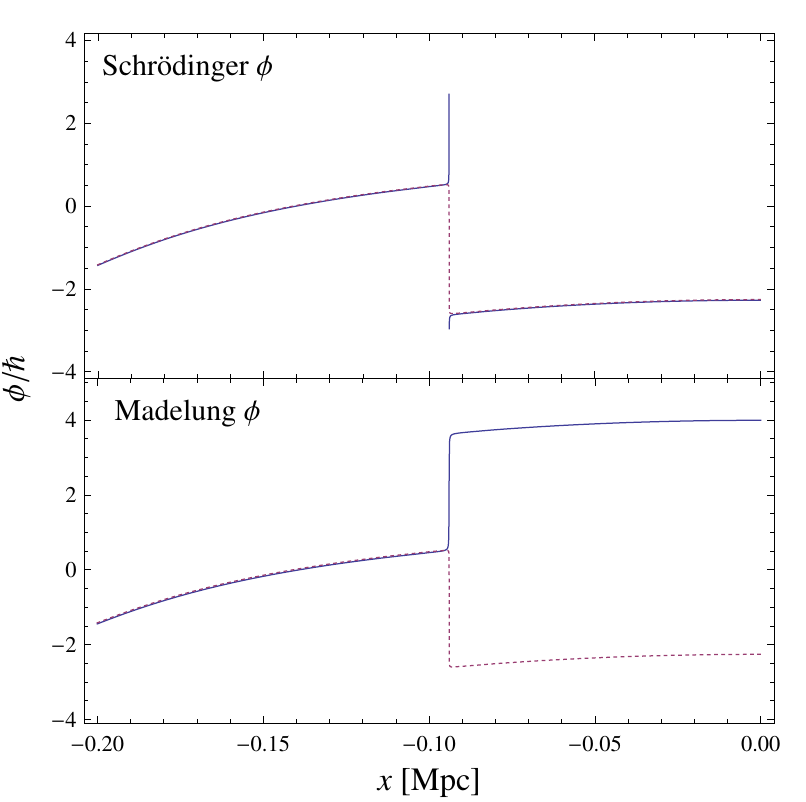}\\
\caption{The first phase jump $\Delta \phi = 2\pi$ occurred around $a_{\phi}\simeq1.07$. }
\label{fig:phasejump}
\end{figure}
\end{center}

We numerically evolve the initial wave function $\psi$ Eqs.\,(\ref{inipsi},\,\ref{delini}) describing a nearly cold and linear CDM overdensity using the SPE \eqref{schrPoissEqFRW}. 
Within the linear regime the phase $\phi$ and amplitude $n$ are basically indistinguishable from $\phi_{\rm d}$ and  $n_{\rm d}$, however once shell-crossing is approached they start to deviate. The occurrence of singularities in $n_{\rm d}$ and phase jumps $\phi$ are the most dramatic differences. 
In Fig.\,\ref{fig:phasejump} we show the phase closely before and after the time of first phase jump $a_{\phi}$, shortly after the time $a=1$, where $n_{\rm d}$ diverges. Shortly before (\textit{full}) and after (\textit{dotted}) $a_{\phi}$, $\phi$ develops very steep gradients (diverging at the the time of phase jump and changing sign). For the wave function $\psi$ this causes no problem since the amplitude $\sqrt{n}$ vanishes when the step becomes infinitely sharp and allows the phase to ``reconnect'' (\textit{upper panel}), while keeping $\psi$ smooth.  For the Madelung representation this causes another problem: at the moment of phase jump, not only $\vnabla \phi$ but also $\dot{\phi}$ diverges on a whole spatial interval (\textit{lower panel}). This second type of divergence is an artifact caused by neglecting the fact that $\phi$ is defined only modulo $2\pi$.

\begin{center}
\begin{figure}[t]
\includegraphics[width=0.44\textwidth]{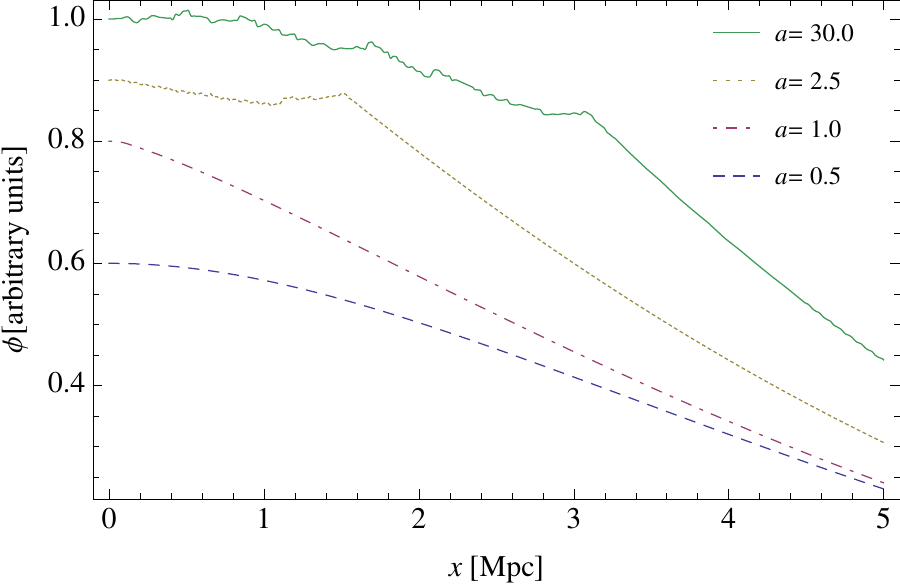}\\
\caption{The phase $\phi$ of the wave function at different times. The wiggly behaviour is characteristic for multi-streaming regions.}
\label{fig:phases}
\end{figure}
\end{center}

\begin{center}
\begin{figure*}[t]
\includegraphics[width=0.49\textwidth]{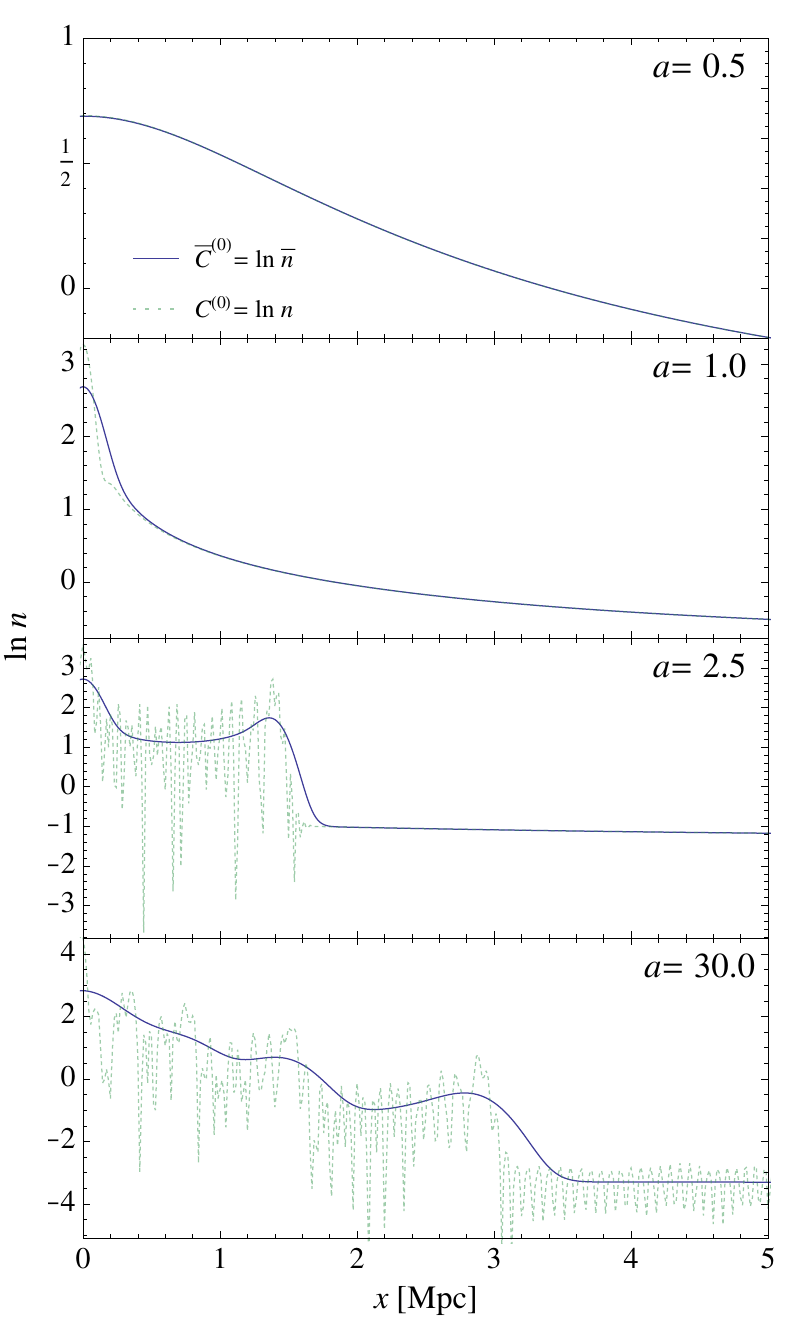}
\includegraphics[width=0.49\textwidth]{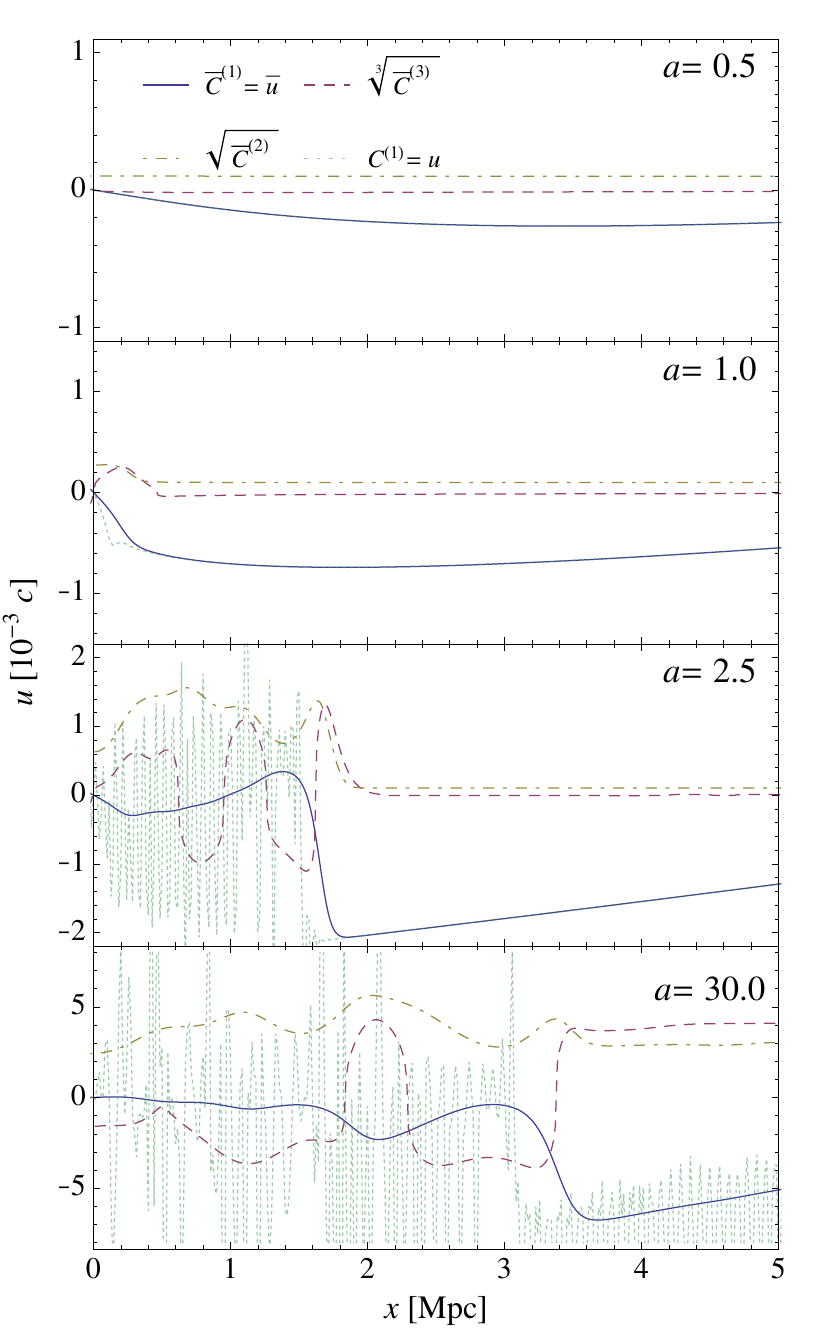}\\
\caption{\textit{left} Number density (\textit{full}) and amplitude squared of wave function (\textit{dotted}).  \textit{right} The first three cumulants and the gradient of phase of the wave function, $\vnabla \phi$. All these quantities are shown at four characteristic times:  the unset of the nonlinear regime around $a=0.5$,  shell-crossing of the dust model at $a=1$, formation of multi-stream regions around the second shell-crossing at $a=2.5$, and  virialization $a=30$. These four times are also shown in Fig.\,\ref{fig:phaseplot}.}
\label{fig:C0C1}
\end{figure*}
\end{center}

At the time $a_{\phi}$ and point $x_\phi$ where the phase develops the sharp step we have $\sqrt{n}=0$. Therefore it makes sense to determine the variance of position and momentum 
 \begin{align}
 \langle x^2\rangle &=\frac{\int_{-x_\phi}^{x_\phi} |\psi|^2 x^2\,d x}{\int_{-x_\phi}^{x_\phi} |\psi|^2\,d x}\quad,\quad
 \langle p^2\rangle=-\hbar^2 \frac{  \int_{-x_\phi}^{x_\phi} \psi^* \Delta \psi \,d x}{\int_{-x_\phi}^{x_\phi} |\psi|^2\,d x}\,.
 \end{align}
Doing the numerical integrals it shows that  $\langle x^2\rangle \langle p^2\rangle \simeq (\hbar/2)^2$, with $\hbar/(2 m) = 10^{-5}\mathrm{Mpc}\,c$ specified for our simulation. The physical interpretation of this result is that the wave function collapsed to its  densest possible state given the initial conditions: a minimum uncertainty wave packet forms within $[-x_\phi, x_\phi]$  at the time $a_\phi$, which expands consequently. We therefore can say that the ScM contains ``shell-crossing without shell-crossing''. This bounce only looks like shell-crossing when coarse-grained over, see App.\,\ref{sec:lagrange}. The result also suggests optimal values for the coarse-graining parameters $\sigp^2 =\langle p^2\rangle$ and $\sigx^2 =\langle x^2\rangle$ of the 1D collapse. We therefore conclude that shell-crossing infinities appearing in $n_{\rm d}$ are now traded for infinities in $\vnabla \phi$, which fortunately do not cause infinities or other pathologies in $\psi$ because $n$ vanishes at those instances and $\psi$ remains smooth.

\begin{center}
\begin{figure}[t!]
\includegraphics[width=0.45\textwidth]{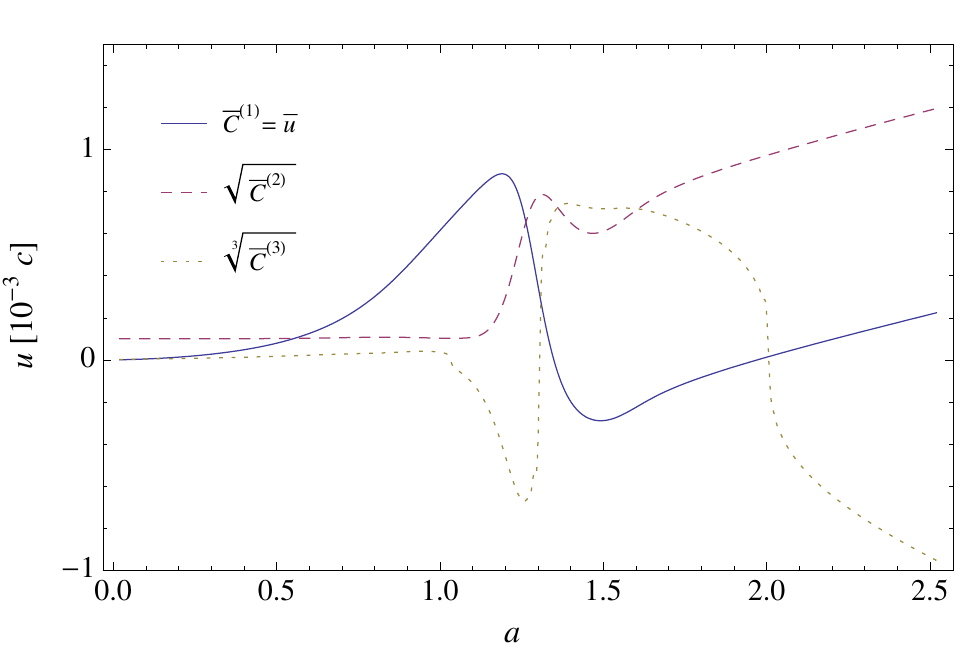}\\
\caption{Comparison between the first three cumulants at the position $x=-0.5\,\mathrm{Mpc}$. They are all equally important after shell-crossing: the hierarchy cannot be truncated.}
\label{fig:C1C2C3}
\end{figure}
\end{center}

\begin{center}
\begin{figure}[t]
\includegraphics[width=0.45\textwidth]{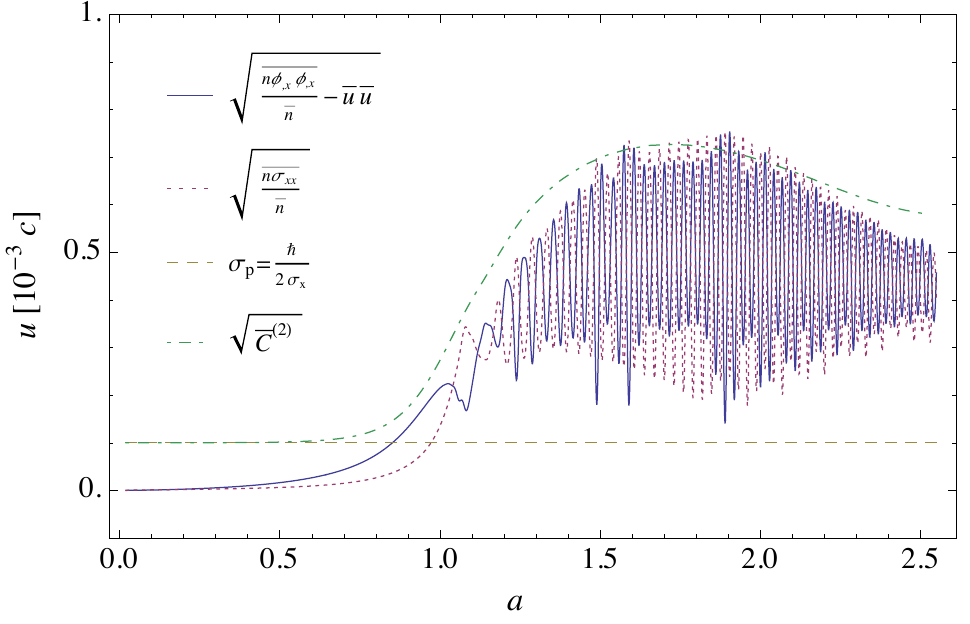}\\
\caption{Comparison between the different parts of the second cumulant at $x=0$. }
\label{fig:C2parts}
\end{figure}
\end{center}

  The wiggly form of the phase, see Fig.\,\ref{fig:phases}, corresponds to large $\vnabla \phi$, which are visible as the strongly oscillating green dotted lines in the right panel of Fig.\,\ref{fig:C0C1}. Because of many phase jumps the amplitude $n$ shows strong spatial oscillations Fig.\,\ref{fig:C0C1}, left.  These oscillations are invisible in the physical quantities of interest: the moments and cumulants of $f_{\rm H}$. We  show the density and the first 3 cumulants in Fig.\,\ref{fig:C0C1} and  Fig.\,\ref{fig:C1C2C3}. They are smooth and physically meaningful. Fig.\,\ref{fig:C1C2C3} also shows that all higher cumulants are switched on at the same time such that the cumulant hierarchy cannot be truncated. In the ScM the two degrees of freedom of $\psi$ store information about all cumulants.
 It is also interesting to note that $\bar{C}^{(2)}$, Eq.\,\eqref{barC2}, can be decomposed into a purely spatial average induced velocity dispersion, a smoothed but microscopic velocity dispersion and a constant part. Most notably, the first two contributions are equally large and show oscillations over time but add up to a smooth sum, see Fig.\,\ref{fig:C2parts}.
Finally let us consider the full phase space dynamics in Fig.\,\ref{fig:phaseplot}. The Husimi distribution $f_{\rm H}$ contains  like $\psi$ the information about all cumulants, but unlike $\psi$, in a form directly related to physical quantities. The most interesting features are the regularity at shell-crossing, the formation of multi-stream regions and the possibility to follow the dynamics until virialization.

Notice that $\bar{C}^{(2)}$ within multi-stream regions remains always positive while $\bar{C}^{(1)}$ basically vanishes.
We therefore checked that the (macroscopic) tensor virial theorem \cite{BT08}, following from the Euler-type equation \eqref{EulercgW} and a steady state assumption (within the virialised object $\bar u = 0$), 
\begin{multline}
 \frac{1}{a^2}\int_{-x_{\rm vir}}^{x_{\rm vir}} \mathrm{d}x\,( \bar{M}^{(2)}_{x x} - \sigp^2 \bar n)=\\ \int_{-x_{\rm vir}}^{x_{\rm vir}} \mathrm{d}x\,x\exp[\tfrac{1}{2}\sigx^2 \Delta]\left( n(x) \partial_x V(x)\right)
\end{multline}
is approximately satisfied for $x_{\rm vir} \simeq 2.8\,\mathrm{Mpc}$ for $a=30$. The $\sigp$-term as well as the boundary terms from integrating by parts are completely negligible. Looking at the right panel of Fig.\,\ref{fig:C0C1} we see that below $x_{\rm vir}$ the macroscopic velocity $\bar u$ is basically zero for $a=30.0$, looking at the left panel we see that the macroscopic density peaks around $x_{\rm vir}$ and drops off afterwards. Note that relaxation is known to take much longer in 1D than 3D \cite{TGK96}.

\section{Prospects} \label{sec:prospects}
For analysing, understanding as well as estimating statistical errors  of observations of LSS one is interested in $n$-point correlation functions of the phase space density. In the ScM these correlation functions are simply related to the $2n$-point correlation functions of the complex scalar $\psi$

\begin{multline*}
\langle f(t, \v{r}_1,\v{p}_1) ...  f(t, \v{r}_{n},\v{p}_{n})\rangle = \\
\Bigg(\prod_{i =1}^n \int \vol{3}{x_i}\! \vol{3}{y_i} K_{\rm H}(\v{r}_i, \v{x}_i, \v{p}_i) K_{\rm H}^*(\v{r}_i, \v{y}_i, \v{p}_i)\Bigg) \times \\
\qquad\langle \psi(t, \v{x}_1) \psi^*(t, \v{y}_{1})... \psi(t, \v{x}_n) \psi^*(t, \v{y}_{n})\rangle\,,\end{multline*}
where $K_{\rm H}$ is the Husimi kernel Eq.\,\eqref{Husimi} and the angle brackets denote know ensemble average over all initial conditions.
This allows the construction of $n$-point redshift space matter and halo correlation functions upon integration over $$\prod_{i=1}^n\delta_{\rm D}\left(\v{s}_i- \v{r}_i- \frac{\v{p}_i \cdot \hat{\v{z}}}{a^2 m H } \hat{\v{z}}\right)\ \vol{3}{p_i}\vol{3}{r_i}\,,$$ where $\hat{\v{z}}$ points along the line of sight and $\v{s}_i$ are the observed positions in redshift space.
As a first step one can study the redshift space  2-point correlation in the case where $\hbar=0$, keeping only $\sigx$ and $\sigp$ \cite{HUK14}. This approach is motivated by the observation that keeping only $\sigx$ results in a resummation in the large-scale parameter of the macroscopic model suggested in \cite{D00,BD05}.

Ultimately we would like to keep $\hbar$, since from our numerical study it is clear that the quantum pressure plays a crucial rule not only in shell-crossing regularization but also within the cumulants, see Fig.\,\ref{fig:C2parts}. Therefore we need a method to calculate the time evolution of $\langle \psi(t, \v{x}_1) ...  \psi^*(t, \v{y}_{n})\rangle$ including $\hbar$ and most desirably in a non-perturbative fashion.
 
 There is a simple Lagrangian and action for $\psi$ from which the SPE follow from the variational principle \cite{AS01}. Therefore one might take the route of \cite{T11} and integrate the nonperturbative renormalisation group flow with time as flow parameter \cite{GKS10}. Another possibility would be to explore the fact that $\hbar$ corresponds to the phase space resolution and thus might be used as a flow parameter with interpretation of Kadanoff's block spin transformation \cite{K66}. 

It might also be possible to interpret the formation of wiggly phases via phase jumps, see Figs.\,\ref{fig:phasejump} and \ref{fig:phases}, as something akin to a phase transition. Halo formation under time evolution would then  correspond to magnetic domain formation or hadronisation in a ferromagnet or quark-gluon plasma, respectively,  under adiabatic cooling.

The ScM could also be connected to effective field theory formulations of LSS formation \cite{B10,PSZ13,CLP13}. Since the ScM is a UV complete theory it might be possible to derive an effective field theory including its parameters.

Another research route could be to look for stationary complex solutions of the SPE\footnote{To our knowledge, so far only real solutions have been studied \cite{AS01,MPT98}.  Fig.\,\ref{fig:phases} however suggests that stationary solutions that result from gravitational collapse are complex.} with the aim of understanding the universality of density profiles of virialised objects. Since ScM allows for virialization it could prove useful in further analytical understanding of violent relaxation \cite{L67,SH92} that leads to universal phase space and density profiles \cite{NFW97, DM14}.

\section{Conclusion}\label{sec:conclusion}
We started with the coupled nonlinear Vlasov-Poisson system \eqref{VlasovPoissonEq} for the phase space distribution function $f$ which is relevant for LSS formation of CDM particles which interact only by means of the gravitational potential. Inspired by the Schr\"odinger method (ScM) proposed in \cite{WK93} for numerical simulations we aimed at employing its ability to describe effects of multi-streaming while including recent studies regarding coarse-grained descriptions of CDM and their implications investigated in \cite{P11,D00}. \\
Following closely \cite{WK93}, we introduced a complex field $\psi$ whose time-evolution is governed by the Schr\"odinger-Poisson equation (SPE) \eqref{schrPoissEqFRW} and constructed the coarse-grained Wigner probability distribution $ \bar f_\W$ according to \eqref{fcgWigner} from this wave function. We derived that the time-evolution of $ \bar f_\W$ is determined by Eq.\,\eqref{cgWignerVlasov} which is in good correspondence to the one governed by the coarse-grained Vlasov equation \eqref{cgVlasovEq}. 
Using a numerical toy example we showed how the ScM is able to regularize shell-crossing singularities and allows to follow the dynamics into the fully nonlinear regime.  Furthermore we showed how higher order cumulants \eqref{momentscgw} like velocity dispersion can be calculated directly from the wave function and that a vorticity is generated by the coarse-graining procedure. 

This means that it suffices to solve the SPE \eqref{schrPoissEqFRW}, express the result obtained for $\psi$ in Madelung form $\sqrt n \exp\left(i\phi/\hbar\right)$, and then simply coarse-grain $n$ and $n \v{\nabla} \phi$ to obtain the physical density $\bar n$ and momentum $m \bar n \bar{\v{u}}$, respectively. In a similar fashion all higher cumulants \eqref{cumulants} following from \eqref{genfuncgW} can be obtained from a solution to SPE \eqref{schrPoissEqFRW}.

We derived the corresponding closed-form fluid-like equations \eqref{FluidcgW} for the smooth density field $\bar n$ and the mass-weighted velocity $\bar{\v{u}}$. This is only possible because the `quantum pressure' term proportional to $\hbar^2$ resolves shell-crossing singularities already on the microscopic level. We showed that solving the macroscopic equations \eqref{FluidcgW}  means closing the hierarchy for the moments of $ \bar f_\W$, without truncating the cumulant hierarchy, thereby proposing a different approach to the closure problem than truncation in terms of cumulants. Indeed, all higher cumulants can be written in terms of of $\bar n$ and $\bar{\v{u}}$.

\section*{Acknowledgement}
We would like to thank Dennis Schimmel for enlightening discussions and the Referee for very helpful suggestions. The work of MK \& CU was supported by the DFG cluster of excellence ``Origin and Structure of the Universe''. The work of TH was supported by TR33 ``The Dark Universe''.
\newcommand{\apjl}{Astrophys. J. Letters}
\newcommand{\apjs}{Astrophys. J. Suppl. Ser.}
\newcommand{\mnras}{Mon. Not. R. Astron. Soc.}
\newcommand{\pasj}{Publ. Astron. Soc. Japan}
\newcommand{\apss}{Astrophys. Space Sci.}
\newcommand{\aap}{Astron. Astrophys.}
\newcommand{\physrep}{Phys. Rep.}
\newcommand{\mpla}{Mod. Phys. Lett. A}
\newcommand{\jcap}{J. Cosmol. Astropart. Phys.}

\bibliography{HusimiVlasovbib}

\newpage
\appendix
\begin{widetext}
\section{Explicit calculation for closing the hierarchy}
\label{closeHierarchy}
As mentioned in \ref{3rdmoment} it can be shown that the evolution equation for the second moment \eqref{consistencycgW} is automatically fulfilled when the coarse-grained fluid equations \eqref{FluidcgW} for density $\bar n$ and mass-weighted  velocity $\bar{\v{u}}$  are satisfied. In order to prove that we perform the following steps:
\begin{enumerate}
\item Start with the time evolution equation for the second moment \eqref{consistencycgW} which involves the third one.
\begin{align} 
\del_t \bar M^{(2)}_{ij}&\stackrel{?}{=} -\frac{1}{a^2  m}\nabla_k \bar M^{(3)}_{ijk} - m \nabla_{(i}\bar V  \exp\left(\sigx^2\overleftarrow{\vnabla}_{\! \!  x}\overrightarrow{\vnabla}_{\! \!  x}\right) (\bar n \bar u_{j)}) + \frac{\sigp^2}{a^2} (\bar n \bar u_{(i})_{,j)} 
\end{align}
\item Insert the explicit expressions for $\bar M^{(2)}$ and $\bar M^{(3)}$ given by \eqref{2momentcgw} and \eqref{3momentcgw}.
\begin{align}
&  \del_t \exp\left( \frac{\sigx^2}{2}\Delta \right)\left[n\phi_{,i}\phi_{,j}+\sigp^2n\delta_{ij} +\frac{\hbar^2}{4} \left(\frac{n_{,i}n_{,j}}{n}-n_{,ij}\right) \right] \\
\notag &\stackrel{?}{=} -\exp\left( \frac{\sigx^2}{2}\Delta\right) \nabla_k\left\{ n\phi_{,i}\phi_{,j}\phi_{,k}+\stackrel{+\text{cyc. perm.}}{\sigp^2 \delta_{ij}n\phi_{,k}}+\frac{\hbar^2}{4} \left[\stackrel{+\text{cyc. perm.}}{\left(\frac{n_{,i}n_{,j}}{n} - n_{,ij}\right)\phi_{,k}}  -n\phi_{,ijk}\right] \right\} - \nabla_{(i}\bar V  \exp(\sigx^2\overleftarrow{\vnabla}_{\! \!  x}\overrightarrow{\vnabla}_{\! \!  x}) (\bar n\bar u_{j)}) + \sigp^2 (\bar n \bar u_{(i})_{,j)} 
\end{align}
\item Express everything in terms of $\bar n$ and $\bar u_{i} = \overline{(n \phi_{,i})}/\bar n$ using the rule for the $D$-symbol \eqref{Dop}.
\begin{align}
& \del_t \left\{ \exp\left[ \frac{\sigx^2}{2}(\Delta-D) \right] \left[\frac{(\bar n \bar u_i)(\bar n \bar u_j)}{\bar n} +\frac{\hbar^2}{4} \left(\frac{\bar n_{,i} \bar n_{,j}}{\bar n}-\bar n_{,ij}\right)\right] +\sigp^2 \bar n \delta_{ij} \right\}\\
\notag &\stackrel{?}{=} -\exp\left[ \frac{\sigx^2}{2}(\Delta-D)\right] \nabla_k \left[ \frac{(\bar n \bar u_i) (\bar n \bar u_j) (\bar n \bar u_k)}{\bar n^2} +\frac{\hbar^2}{4} \stackrel{+\text{cyc. perm.}}{\left[\left(\frac{\bar n_{,i} \bar n_{,j}}{\bar n} - \bar n_{,ij}\right)\frac{\bar n\bar u_k}{\bar n} - \frac{1}{3}\bar n\left(\frac{\bar n \bar u _i}{\bar n}\right)_{,jk}\right]} \right]\\
\notag&\quad -\sigp^2  \nabla_k\stackrel{+\text{cyc. perm.}}{\left( \delta_{ij}\bar n\bar u_k \right)}  - \nabla_{(i}\bar V  \exp(\sigx^2\overleftarrow{\vnabla}_{\! \!  x}\overrightarrow{\vnabla}_{\! \!  x}) (\bar n\bar u_{j)}) + \sigp^2 (\bar n \bar u_{(i})_{,j)} 
\end{align}
\item Pull the time-derivative through the smoothing operator and apply the product rule to re-express the terms.
\begin{align}
&\exp\left[ \frac{\sigx^2}{2}(\Delta-D) \right]  \left\{ \frac{\del_t (\bar n \bar u_{(i})(\bar n\bar u_{j)})}{\bar n} - \frac{(\bar n\bar u_i)(\bar n\bar u_j)\del_t \bar n}{\bar n^2} + \frac{\hbar^2}{4} \left(\frac{\del_t \bar n_{,(i}\bar n_{,j)}}{\bar n}-\frac{\del_t \bar n\ \bar n_{,i}\bar n_{,j}}{\bar n^2}-\del_t \bar n_{,ij}\right) \right\} +\sigp^2 \del_t \bar n \delta_{ij}\\
\notag &\stackrel{?}{=} \exp\left[ \frac{\sigx^2}{2}(\Delta-D) \right] \left\{ -\nabla_k \left(\frac{\bar n\bar u_i\bar n\bar u_k}{\bar n}\right)\frac{\bar n\bar u_j}{\bar n} - \frac{\bar n\bar u_i\bar n\bar u_k}{\bar n}\nabla_k \left(\frac{\bar n\bar u_j}{\bar n}\right)  - \nabla_{(i}\bar V  (\bar n\bar u_{j)}) - \frac{\hbar^2}{4} \nabla_k\stackrel{+\text{cyc. perm.}}{\left[\left(\frac{\bar n_{,i}\bar n_{,j}}{\bar n} - \bar n_{,ij}\right)\frac{\bar n\bar u_k}{\bar n}  - \frac{1}{3}\bar n\left(\frac{\bar n \bar u _i}{\bar n}\right)_{,jk}\right]} \right\}  -\sigp^2 \nabla_k(\bar n\bar u_k) \delta_{ij}
\end{align}
\item Employ the fluid equations \eqref{FluidcgW} to carry out the time derivatives $\del_t(\bar n) $ and $\del_t(\bar n \bar u_i)$.
\begin{align}
&\exp\left[ \frac{\sigx^2}{2}(\Delta-D) \right]  \left\{-\exp\left[ \frac{\sigx^2}{2}(\Delta-D) \right] \left[ \nabla_k \left(\frac{\bar n \bar u_k\bar n\bar u_{(i}}{\bar n}\right) + \nabla_{(i} \bar V  \bar n + \frac{\hbar^2}{4} \nabla_k\left(\frac{\bar n_{,k}\bar n_{,(i}}{\bar n} - \bar n_{,k(i}\right)\right]\frac{\bar n \bar u_{j)}}{\bar n} +\frac{\bar n\bar u_i\bar n\bar u_j(\bar n \bar u_k)_{,k}}{\bar n^2}\right.\\
\notag&\qquad \qquad \qquad \qquad \left.  - \frac{\hbar^2}{4} \left(\frac{(\bar n\bar u_k)_{,k(i}\bar n_{,j)}}{\bar n}-\frac{(\bar n\bar u_k)_{,k}\bar n_{,i}\bar n_{,j}}{\bar n^2}-(\bar n\bar u_k)_{,ijk}\right)  \right\} \\
\notag &\stackrel{?}{=} \exp\left[ \frac{\sigx^2}{2}(\Delta-D) \right] \left\{ -\nabla_k \left(\frac{\bar n\bar u_i\bar n\bar u_k}{\bar n}\right)\frac{\bar n\bar u_j}{\bar n} - \frac{\bar n\bar u_i\bar n\bar u_k}{\bar n}\nabla_k \left(\frac{\bar n\bar u_j}{\bar n}\right)- \nabla_{(i}\bar V  (\bar n\bar u_{j)})- \frac{\hbar^2}{4} \nabla_k\stackrel{+\text{cyc. perm.}}{\left[\left(\frac{\bar n_{,i}\bar n_{,j}}{\bar n} - \bar n_{,ij}\right)\frac{\bar n\bar u_k}{\bar n}  - \frac{1}{3}\bar n\left(\frac{\bar n \bar u _i}{\bar n}\right)_{,jk}\right]} \right\}  
\end{align}
\item Combine the different $D$-symbols acting successively on the terms to yield an overall $D$-symbol according to
$$\exp\left[ \tfrac{1}{2} \sigx^2(\Delta-D_{ABC}) \right]\left(\bar A \bar B \bar C\right) = \exp\left[ \tfrac{1}{2} \sigx^2(\Delta-D_{A(BC)}) \right]\left[\bar A\exp\left[ \tfrac{1}{2} \sigx^2(\Delta-D_{BC}) \right] (\bar B \bar C)\right] \,. $$
This is possible since the action of the $D$-symbol depends on the product structure it is acting on. 
\begin{align}
& \exp\left[ \frac{\sigx^2}{2}(\Delta-D) \right]  \left\{ \frac{\hbar^2}{4} \left[ \nabla_k\left(\frac{\bar n_{,k}\bar n_{,(i}}{\bar n} - \bar n_{,k(i}\right)\frac{\bar n\bar u_{j)}}{\bar n} + \frac{(\bar n\bar u_k)_{,k(i}\bar n_{,j)}}{\bar n}-\frac{(\bar n\bar u_k)_{,k}\bar n_{,i}\bar n_{,j}}{\bar n^2}-(\bar n\bar u_k)_{,ijk} \right] \right\} \\
\notag &\stackrel{\checkmark}{=}   \exp\left[ \frac{\sigx^2}{2}(\Delta-D) \right]  \left\{ \frac{\hbar^2}{4} \nabla_k\stackrel{+\text{cyc. perm.}}{\left[\left(\frac{\bar n_{,i}\bar n_{,j}}{\bar n} - \bar n_{,ij}\right)\frac{\bar n\bar u_k }{\bar n} - \frac{1}{3}\bar n\left(\frac{\bar n \bar u _i}{\bar n}\right)_{,jk}\right]} \right\}
\end{align}
One has to note that equality is only established once we make use of the constraint Eq.\,\eqref{Husimiconstr}.
\end{enumerate}
\end{widetext}

\section{Lagrangian formulation}
\FloatBarrier
\label{sec:lagrange}
\begin{center}
\begin{figure}[t!]
\includegraphics[width=0.48\textwidth]{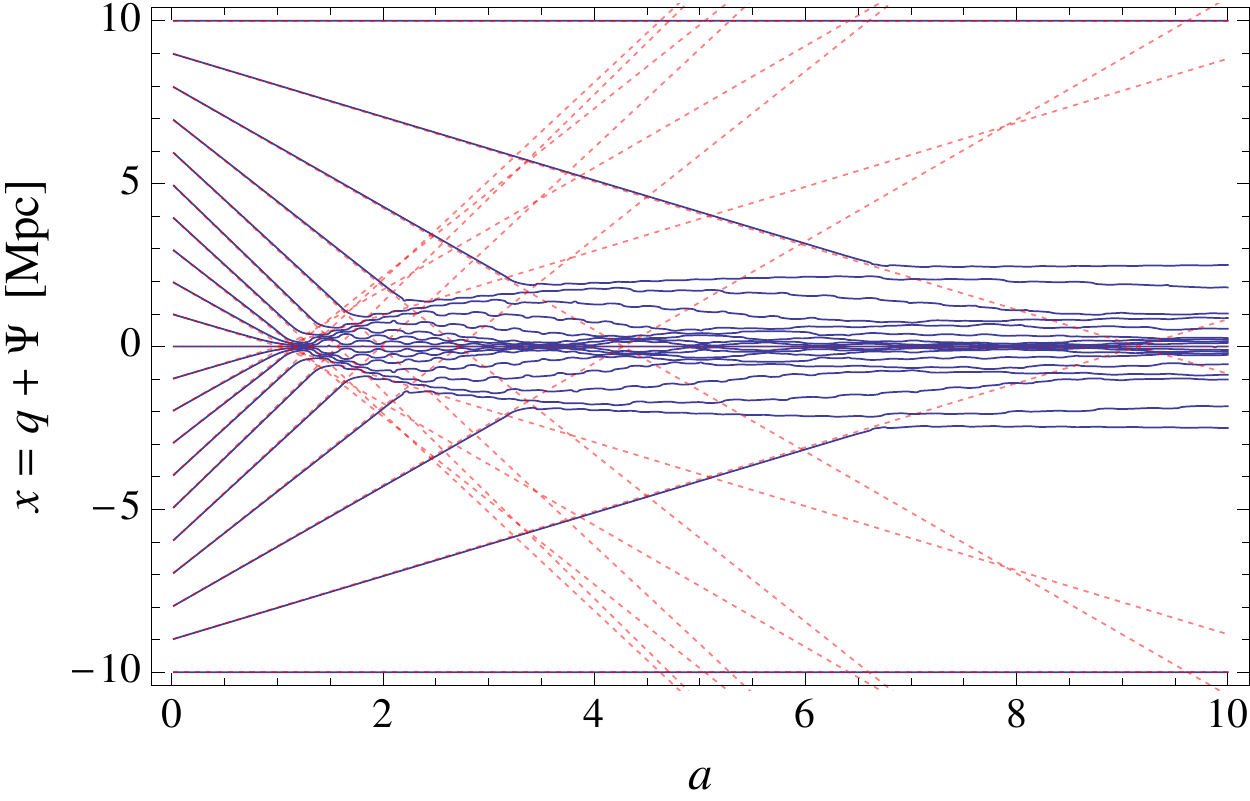}\\
\caption{\textit{red dotted} Zel'dovich trajectories Eq.\,\eqref{ZeldoPancake}, \textit{blue} Bohmian trajectories Eq.\,\eqref{LagrEq1D}. }
\label{fig:trajectories}
\end{figure}
\end{center}
We follow \cite{RB12} to rewrite the fluid-like system Eqs.\,\eqref{Madelungfluideq} formulated in terms $n$ and $\vnabla \phi$ evaluated at the Eulerian position $\v{x}$, into a Lagrangian system in which the sole dynamical variable is the displacement field $\v{\varPsi}$, that maps between $\v{x}$ and the Lagrangian (or initial coordinate of a fluid element) $\v{q}$. Since the continuity and Euler equation Eqs.\,\eqref{Fluideq} are unchanged apart from the added quantum potential $Q$ in Eq.\,\eqref{EulerMadelung} the analogue of Eq.\,2.31 in \cite{RB12} is
\begin{align} \label{LagrEq3D}
 \left[(1+ \varPsi_{l,l}) \delta_{ij}- \varPsi_{i,j} +\varPsi^c_{i,j}) \right] \varPsi_{i,j}'' =  \qquad \qquad \qquad \qquad \\ \qquad\alpha(\eta) (J^F -1) + J^F \frac{\hbar^2}{4 m^2} \Delta_x\left( \frac{\Delta_x [(J^F)^{-1/2}]}{(J^F)^{-1/2}} \right) \notag \,,
\end{align}
which can be obtained by solving the continuity equation with $1+\delta = 1/J^F$, where $J^F = \det (F_{ij}) = \det (\delta_{i,j} + \varPsi_{i,j})$ and $F_{ij} = \partial x^i /\partial q^j$ is the Jacobian relating $\v{x}$ and $\v{q}$ and with $\vnabla_{\! \!  x} \phi/m = \v{\varPsi}'$, where a prime denotes a derivative wrt to superconformal time $\eta$ related to cosmic time $t$ via $dt= a^2 d\eta $.
In eq.\,\eqref{LagrEq3D} the Laplacians are with respect to $\v{x}$, rather than $\v{q}$ and have therefore to be rewritten in terms of $\v{q}$ using the Jacobian $F_{ij}$. The equation is supplemented by a constraint equation $F_{i,n}\epsilon_{njk}F_{l,j}F'_{l,k}=0$ that follows from $\v{\nabla}_x \times \v{u} =0$. 
If the density and velocity distribution depend only on $\v{x}=(x,0,0)$, (and therefore $\v{q}=(q,0,0)$), the above system can be written, using $\epsilon_{qqq} =0$ and $\varPsi_i=: \varPsi \delta_{i q}$ and $J^F = 1+ \varPsi_{,q}$ as
\begin{equation} \label{LagrEq1D}
\varPsi '' = \alpha(\eta) \varPsi + \frac{\hbar^2}{2 m^2} \left( \frac{10 (\varPsi_{,qq})^3}{(1+\varPsi_{,q})^6} -8 \frac{\varPsi_{,qq} \varPsi_{,qqq}}{(1+\varPsi_{,q})^5}+ \frac{ \varPsi_{,qqqq}}{(1+\varPsi_{,q})^4}\right) \,,
\end{equation}
where $\alpha(\eta) = 4 \pi G a \rho_0$. Note that compared to the 3D case \eqref{LagrEq3D}, we were able to integrate already once over $q$ in order to obtain \eqref{LagrEq1D}. 
\vspace{0.5cm}
\begin{center}
\begin{figure}[t!]
\includegraphics[width=0.48\textwidth]{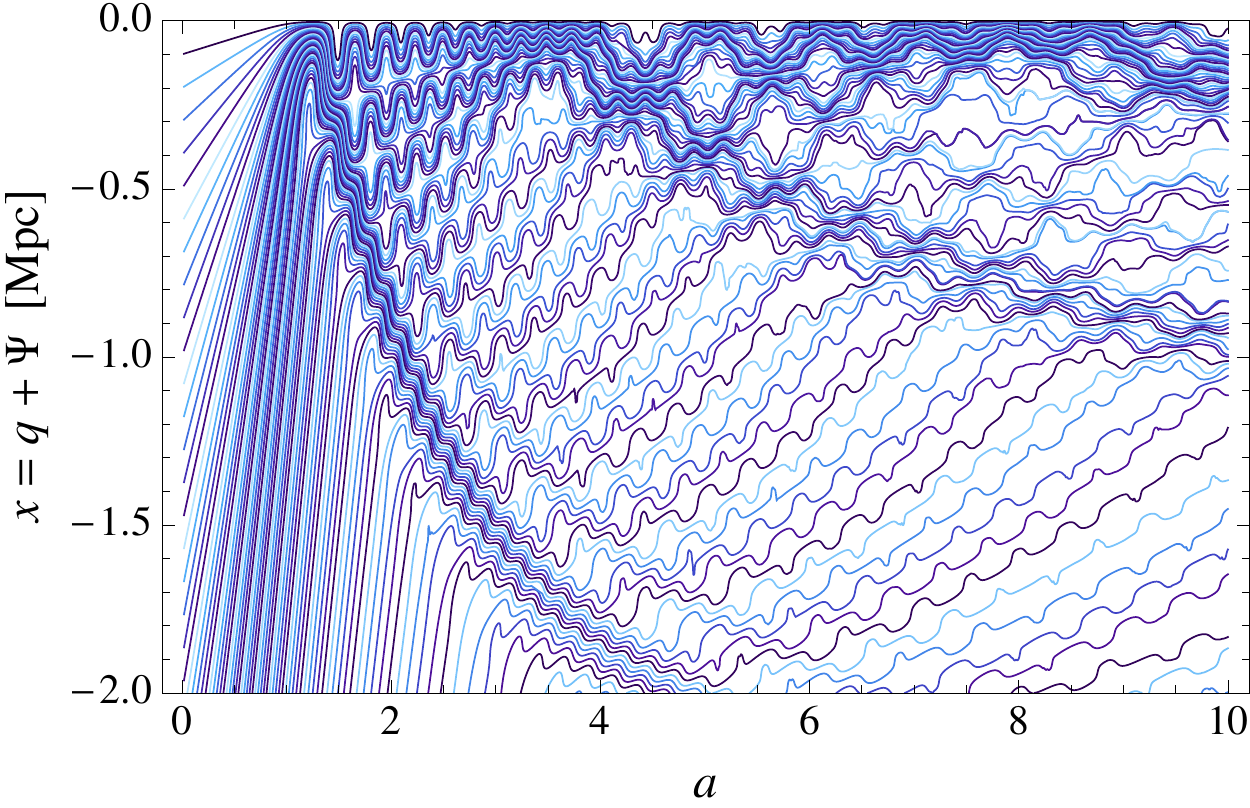}\\
\caption{Detailed view of the Bohmian trajectories, Eq.\,\eqref{LagrEq1D}.}
\label{fig:trajectoriestrjactoriesDetail}
\end{figure}
\end{center}
In the case of $\hbar =0$, we recover the case of dust
\begin{align} \label{ZeldoPancake}
 \varPsi_{\rm d} '' = \alpha(\eta) \varPsi_{\rm d}\,,
 \end{align}
  whose exact solution is the Zel'dovich approximation $\varPsi_{,q}(\v{q},a) = -D(a) \delta_{\rm lin} (\v{x} = \v{q})$, where $\delta_{\rm lin} (\v{x})$ is the initial condition Eulerian density field (which is assumed to vanish at $a=0$) linearly extrapolated to $a=1$ using the linear growth $D(a)$. The red dashed lines in Fig.\,\ref{fig:phaseplot} are points $(q+\varPsi, \varPsi')$, parametrized by $q$ and can be extended after shell-crossing. Unfortunately, this continuation does not behave as CDM and the trajectories continue on their straight lines indefinitely, see red lines in Fig.\,\ref{fig:trajectories}. Including the $\hbar$-terms, a separation ansatz does not work anymore and we do not expect to find an exact solution of \eqref{LagrEq1D}, see Figs.\,\ref{fig:trajectories} and \ref{fig:trajectoriestrjactoriesDetail} for the complicated dynamics of $\varPsi$ for the case of initial conditions studied in Sec.\,\ref{sec:numerics}.  Under a coarse-grained view the Bohmian and collisionless CDM trajectories would turn into network that is indistinguishable. On a microscopic level though, they are very different, see Fig.\,\ref{fig:trajectoriestrjactoriesDetail}. Although the phase space density $f_{\rm H}$ behaves as if shell-crossings and multi-stream regions form, the phase $\phi$ of the wave function $\psi$ is single-valued and therefore the trajectories $\v{q}+\v{\varPsi}$ never intersect. The intricate behaviour  of $\v{\varPsi}$ \textit{emulates} multi-streaming.  Given the Bohmian trajectories $\v{\varPsi}(\v{q},a)$ one can recover $n(\v{x},a)$ and $\phi(\v{x},a)$ via 
\begin{align} \label{Zeldosol}
n(x,a) &= \frac{1}{1+ \varPsi_{,q} (q,\, a)} \Big|_{q= q (x,a)}\\
\partial_x \phi(x,a)/m &=  \varPsi'  (q,\, a) \Big|_{q= q (x,a)}\,,
\end{align}
where the $q$-dependent expressions are converted into $x$-depend ones via inversion of $x = q + \varPsi(q,a)$.
The Lagrangian formulation Eq.\,\eqref{LagrEq3D} of the Madelung representation, Eq.\,\eqref{MadelungFluid}  suffers from the same singularities as the Euler-type equation Eq.\,\eqref{EulerMadelung}; at the isolated space-time points where the phase $\phi$ jumps about $2 \pi$, the velocity $\v{\nabla} \phi$ and therefore $\dot{\v{\varPsi}}$ diverge and change sign. Figs.\,\ref{fig:trajectories} and \ref{fig:trajectoriestrjactoriesDetail} were constructed from the solution of the Schr\"odinger-Poisson equation \eqref{schrPoissEqFRW} and not from Eq.\,\eqref{LagrEq1D}.

\end{document}